\documentclass[preprint,authoryear,12pt]{elsarticle}
\usepackage{epsfig}
\usepackage{amssymb}
\usepackage[dvips]{color}
\usepackage[ps2pdf,%
a4paper=true,%
breaklinks=true,%
colorlinks=true,%
pdfauthor={First Author et al.},%
pdftitle={Template for manuscripts in Advances in Space Research}%
]{hyperref}
\usepackage{natbib}
\journal{Advances in Space Research}

\definecolor{green2} {rgb}{0.0,0.5,0.1}

\def\Rsun{R_\odot}

\begin{document}

\begin{frontmatter}

\title{Seeing The Solar Corona in Three Dimensions}

\author[address1,address2]{Alberto M. V\'asquez\corref{cor}}
\address[address1]{Instituto de Astronom\'\i a y F\'\i sica del Espacio (IAFE, CONICET-UBA), CC 67 - Suc 28, (C1428ZAA) Ciudad de Buenos Aires, Argentina.}
\address[address2]{Facultad de Ciencias Exactas y Naturales (FCEN), University of Buenos Aires (UBA), Ciudad de Buenos Aires, Argentina.}
\cortext[cor]{Corresponding author}
\ead{albert@iafe.uba.ar}

\begin{abstract}
The large availability and rich spectral coverage of today's observational data of the solar corona, and the high spatial and temporal resolution of many instruments, has enabled the evolution of three-dimensional (3D) physical models to a great level of detail. However, the 3D information provided by the data is rather limited as every instrument observes from a single angle of vision, or two at the most in the case of the STEREO mission. Two powerful available observational techniques to infer detailed 3D information of the solar corona from empirical data are stereoscopy and tomography. In particular, the technique known as \emph{differential emission measure tomography} (DEMT) allows determination of the 3D distribution of the coronal electron density and temperature in the inner corona. This paper summarizes the main technical aspects of DEMT, reviews all published work based on it, and comments its future development and applications.
\end{abstract}


\end{frontmatter}

\parindent=0.5 cm

\section{Introduction}\label{S_intro}

The solar corona can be observed in the white light, EUV, X, and radio wavelengths. Being the corona optically thin in these spectral ranges, its images are two-dimensional (2D) projections of the 3D emitting structure. Detailed knowledge of the 3D distribution of the fundamental plasma parameters of the solar corona ($\mathbf{B}$, $N_e$, $T_e$) is highly desirable to advance its modeling. Stereoscopy and tomography are powerful observational techniques of the corona, allowing to infer quantitative 3D information of it. An excellent introduction to solar stereoscopy can be found in \citet{inhester_2006}. A recent general review on both techniques covering all the spectral ranges listed above can be found in \citet{aschwanden_2011}, with a strong focus on stereoscopy. In this review we specifically focus on \emph{differential emission measure tomography} (DEMT) in a more extensive fashion, updating on all published work in the field at the moment of writing this article.

\citet{minnaert_1930} originally developed the scattering theory of the photospheric white light (WL) by the free electrons of the corona, that allows to infer the 3D distribution of the electron density of the corona from WL images. \citet{vandehulst_1950} were the first to perform a global corona reconstruction using eclipse images and assuming full azimuthal axi-symmetry, an assumption firstly relaxed by \citet{leblanc_1970}. It was \citet{altschuler_1972} who developed the first actual solar rotational tomography (SRT) using coronagraph data. A good review on WL SRT can be found in \citet{frazin_2000} and \citet{frazin_2002}, who developed a robust, regularized, positive method for tomographic inversion of the coronal density from time series of WL images. Later on, \citet{frazin_2005} first introduced the concept of DEMT, a technique which uses time series of EUV images to determine the 3D distribution of the coronal local-DEM (or LDEM).

DEMT was developed by \citet{frazin_2009}, and firstly applied by \citet{vasquez_2009} to study the 3D structure of coronal prominence cavities. The technique consists of two steps. In a first step the tomographic inversion of time series of full-sun EUV images is performed, to find the 3D distribution of the EUV emissivity in each filter band of the telescope. In a second step the emissivities found for all bands in any given coronal location are used as a constraint to infer the LDEM. Finally, moments of the LDEM are taken, as a result of which global 3D maps of the coronal electron density and temperature are produced. 

In this review we summarize the main aspects and applications of the DEMT technique. Sections \ref{FBE} and \ref{LDEM} describe and illustrate the two steps of DEMT, section \ref{results} is a review of published results using the technique, and section \ref{conclusions} summarizes its main characteristics and future prospects for its development and application.

\section{The Tomographic Model of the Corona}\label{FBE}

To perform the EUV tomography, the inner corona volume in the {height range 1.00-1.25 $\Rsun$ is discretized on a 25$\times$90$\times$180 (radial $\times$ latitudinal $\times$ longitudinal) spherical grid. Due to optical depth issues (analyzed in detail in \citet{frazin_2009}) and EUV signal-to noise levels (which depend on the particular filter), the results are reliable typically in the height range from 1.03 to 1.20 $\Rsun$}. 

For each filter band of the EUV telescope separately, time series of full-sun EUV coronal images covering a complete solar rotation are used to find the 3D distribution of an emissivity-type quantity known as the \emph{filter band emissivity} (FBE). The FBE of each EUV filter is the integral over wavelength of the coronal spectral emissivity multiplied by the passband of the filter. The intensity in each pixel is a line-of-sight integral of the FBE. The intensities of all pixels of all images can be arranged in a single very large column vector, as well as the FBE in every cell (or voxel) of the tomographic grid. In this way, both vectors are linearly related through a very large non-square sparse projection matrix, that depends on the geometry of the observations. Both the projection matrix and the pixel intensity vector are known, and the problem is to find the FBE vector. This poses a non-invertible linear problem for each band separately, which is the tomographic problem. 

In the case of both the instrument \emph{Extreme ultraviolet Imaging Telescope} (EIT) on board the \emph{Solar and Heliospheric Observatory} (SoHO), and the instrument \emph{Extreme Ultra Violet Imager} (EUVI) on board the \emph{Solar TErrestrial RElations Observatory} (STEREO), the number of EUV bands that can be used for DEMT is 3. In the case of the \emph{Atmospheric Imaging Assembly} (AIA) on board the \emph{Solar Dynamics Observatory} (SDO) the number of bands is increased to 6.

The 3D distribution of the FBE is found by solving a global optimization problem, and the FBE distribution that best reproduces all intensities is determined. A thorough technical explanation of all aspects of the inversion can be found in \citet{frazin_2009}, and discussions on the uncertainties involved can be found in \citet{vasquez_2009, vasquez_2010, vasquez_2011}. 

\begin{figure}[!ht]
\label{tomography}
\begin{center}
\includegraphics[width=\linewidth]{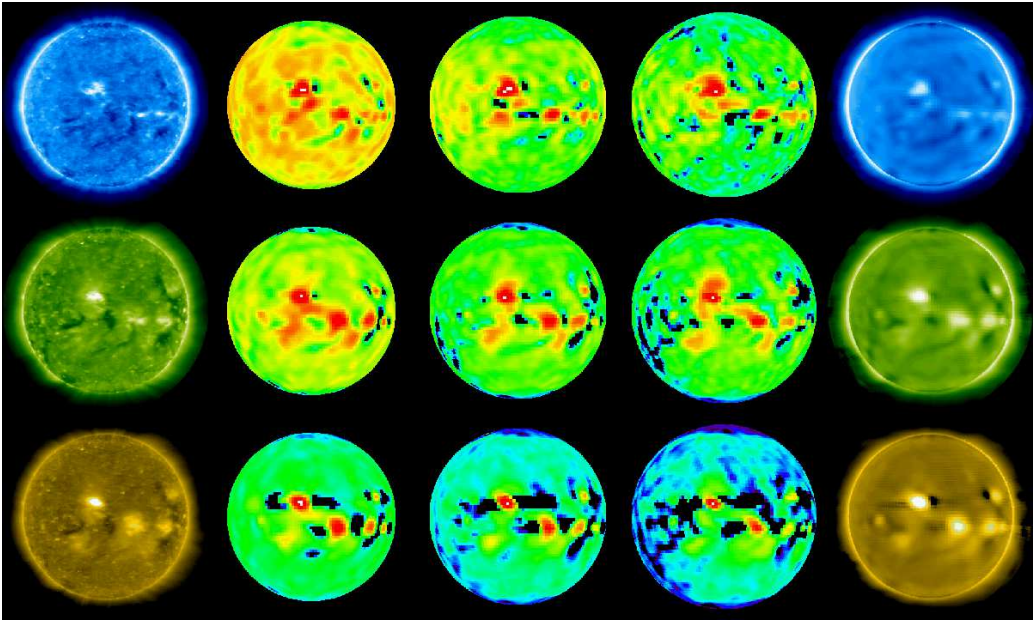}
\end{center}
\caption{A summary of the EUV tomography.  First column shows (from top to bottom) EUVI-A images in the 171, 195 and 284 \AA\ bands taken near 08:00 on 28 April 2008. For each band, the second through fourth columns show projected spherical cuts of the tomographic FBE at 1.035, 1.085, and 1.135 $\Rsun$, respectively. The last column shows the respective synthetic images calculated by integrating the tomographic models along the line-of-sight. The black streaks seen in the reconstructions near some of the active regions are artifacts caused by the Sun's temporal variability. From \citet{vasquez_2009}.}
\end{figure}

For the EUVI instrument, Figure \ref{tomography} shows a summary of the EUV tomography step. The first column shows (from top to bottom) EUVI images in the 171, 195 and 284 \AA\ bands. These images are just one sample from the time series actually used to perform the tomography. For each band, the second through fourth columns show projected spherical cuts of the tomographic FBE at 1.035, 1.085, and 1.135 $\Rsun$, respectively. The last column shows the respective synthetic images calculated by integrating the tomographic models along the line-of-sight. The black streaks seen in the reconstructions near some of the active regions are artifacts caused by the Sun's temporal variability.

\begin{figure}[!ht]
\label{CompareImages}
\begin{center}
\includegraphics[height=0.28\linewidth]{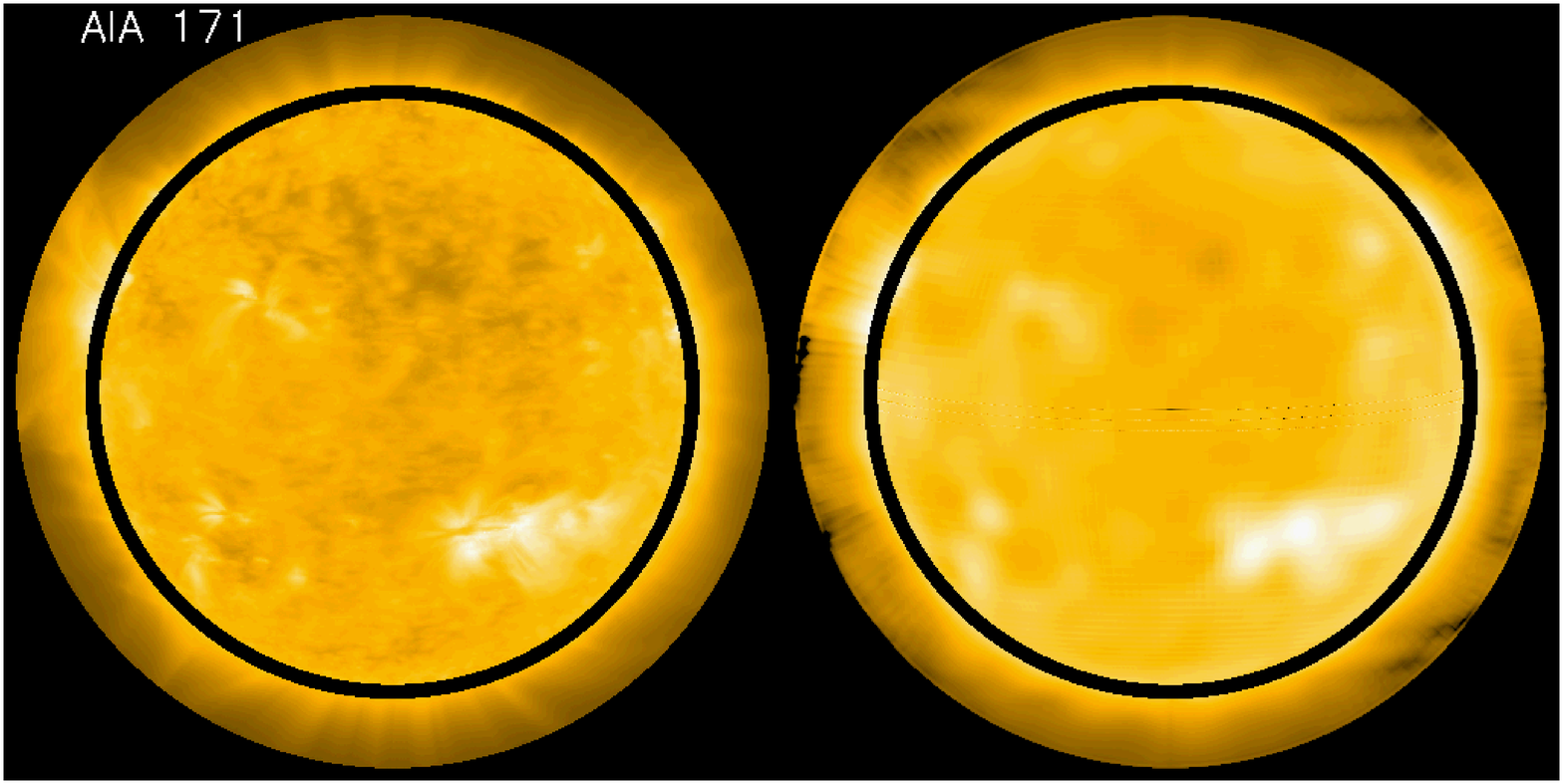}
\includegraphics[height=0.28\linewidth]{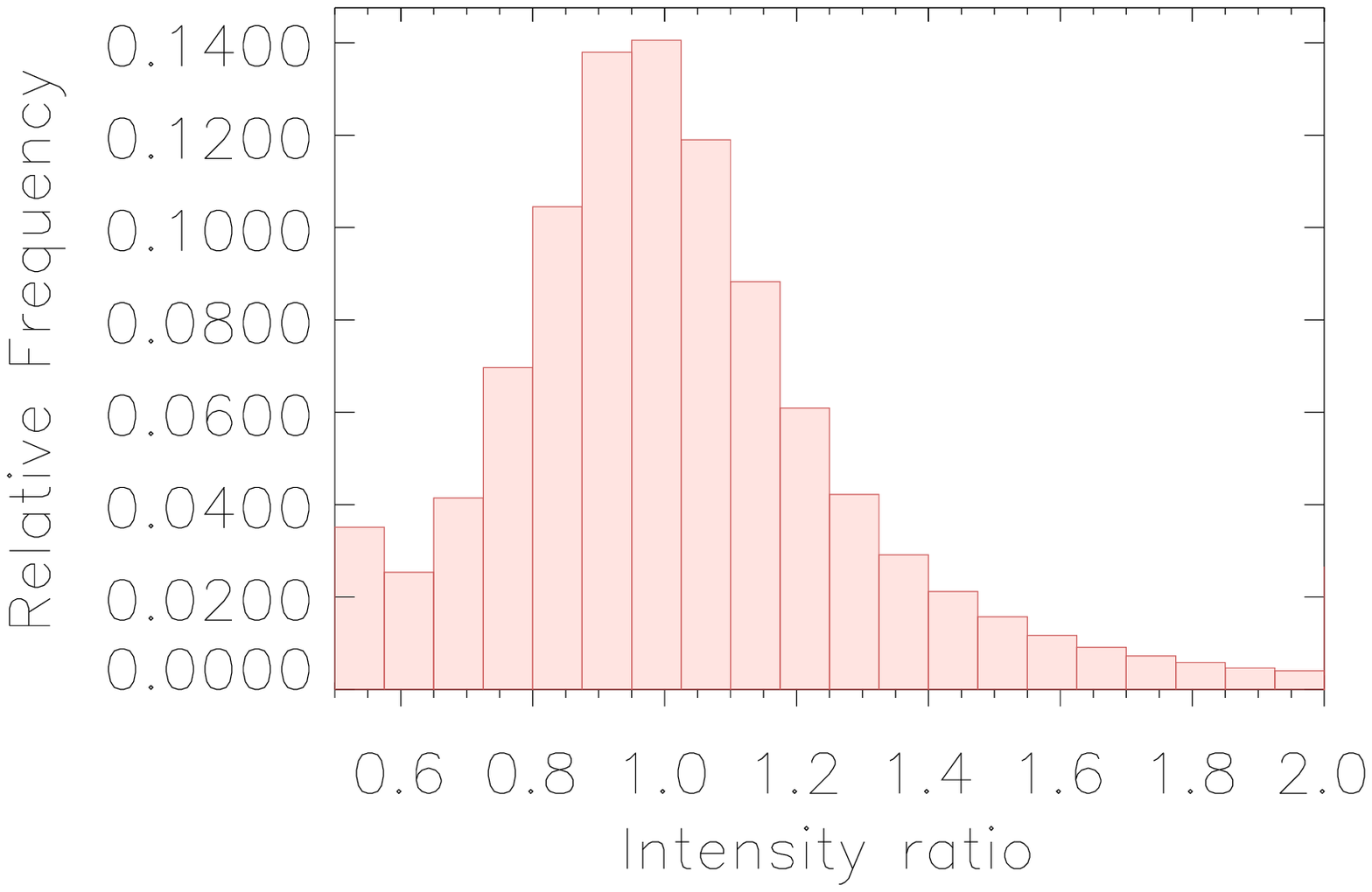}\\
\includegraphics[height=0.28\linewidth]{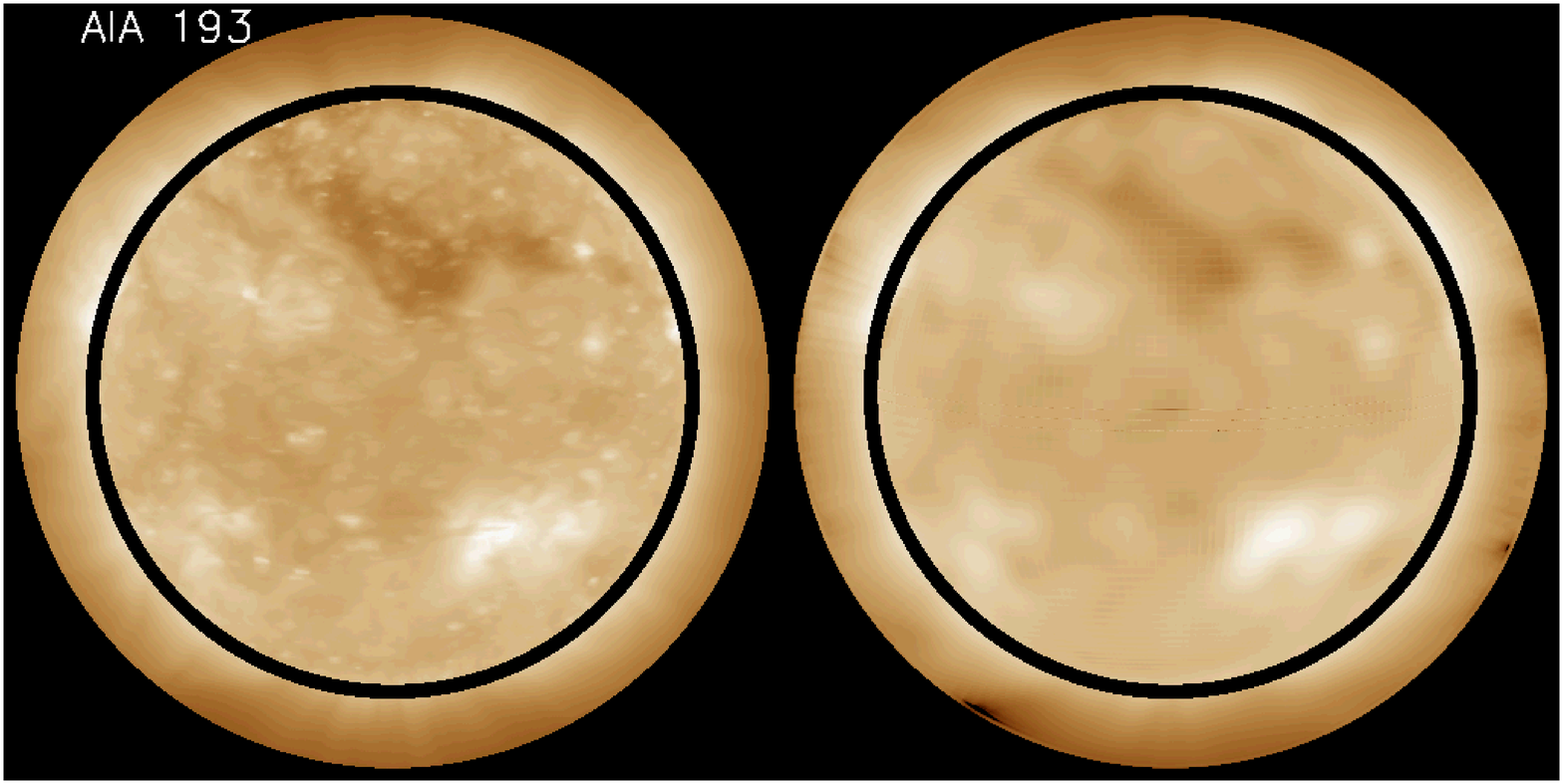}
\includegraphics[height=0.28\linewidth]{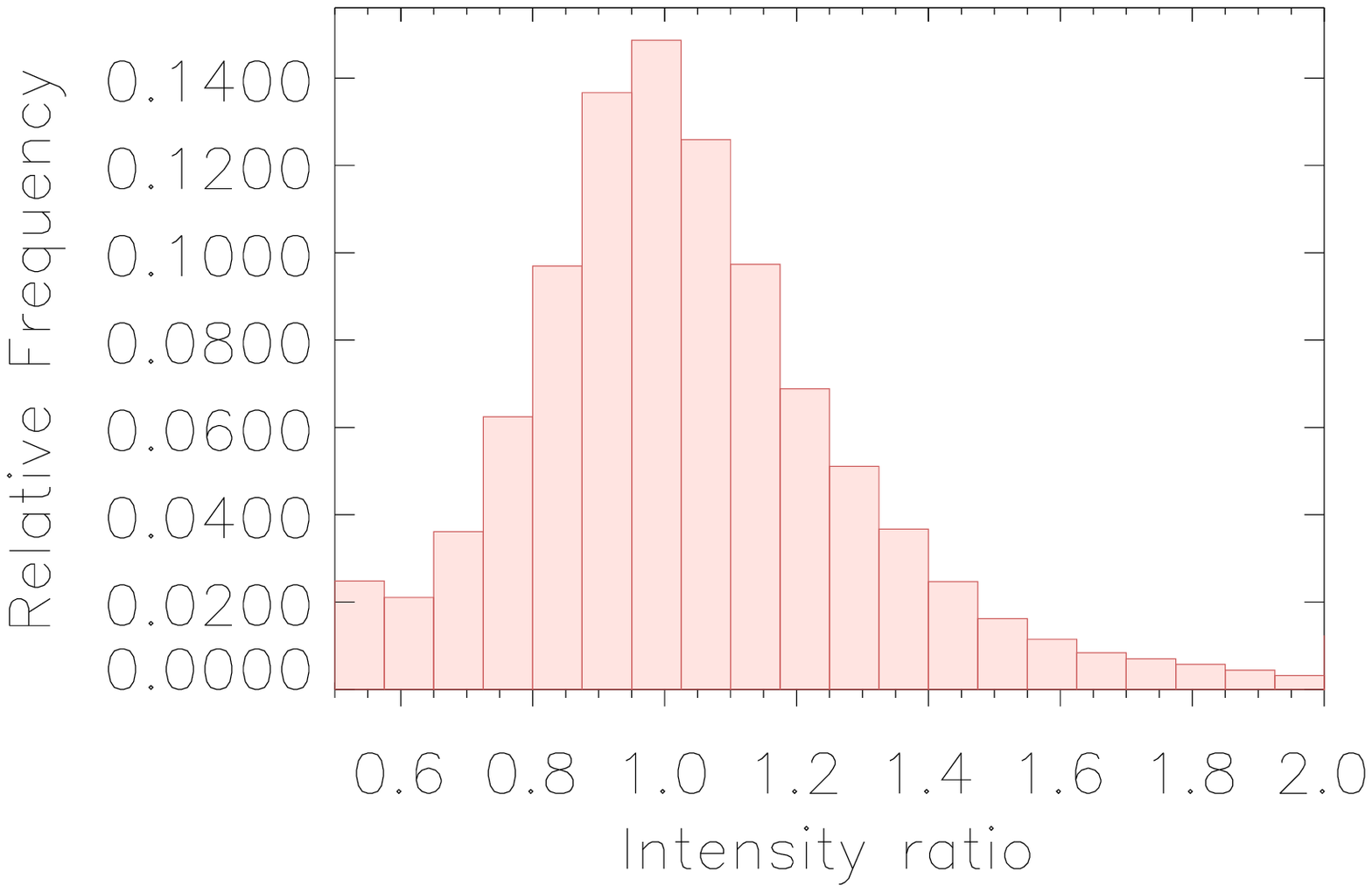}\\
\includegraphics[height=0.28\linewidth]{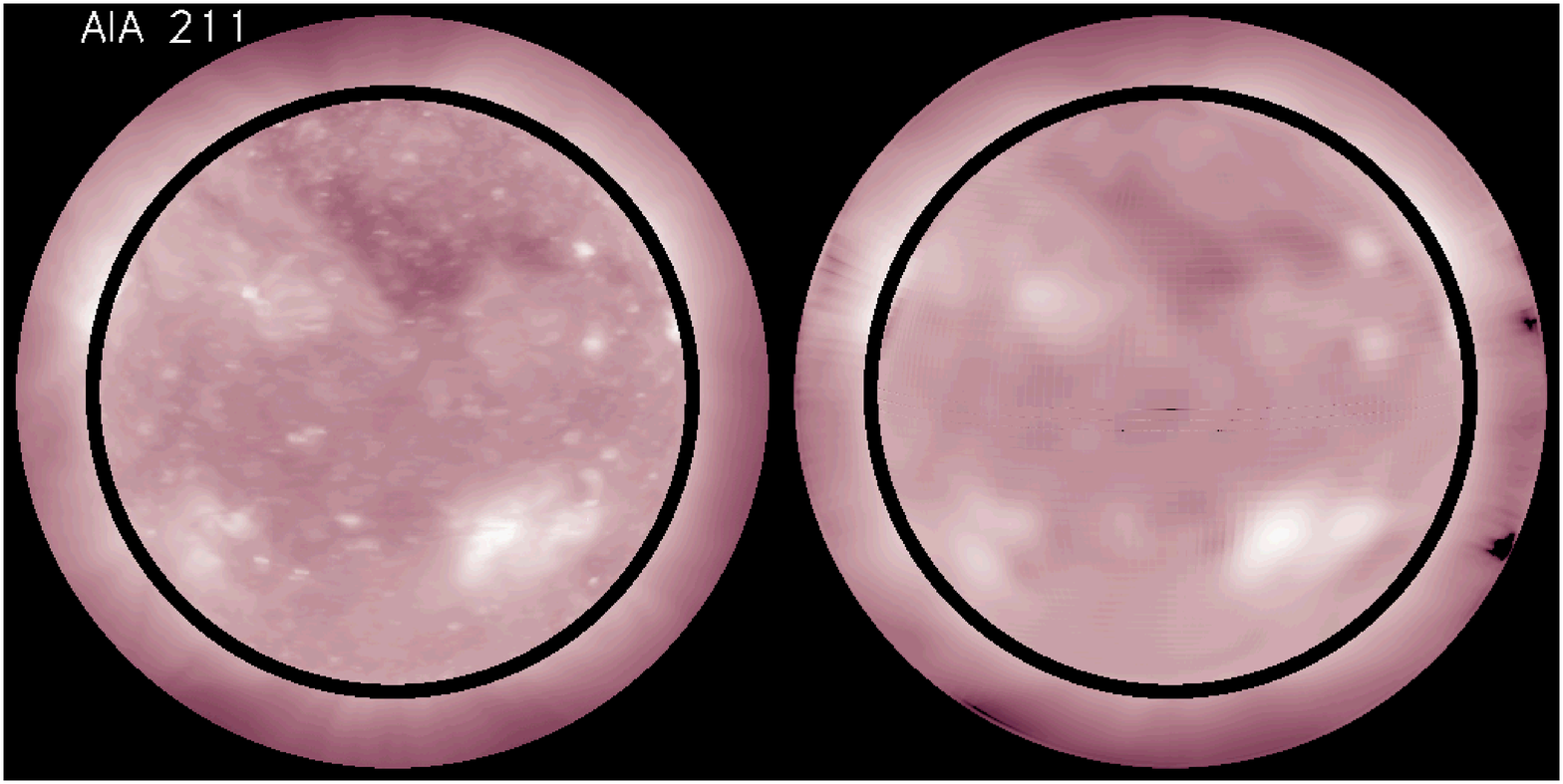}
\includegraphics[height=0.28\linewidth]{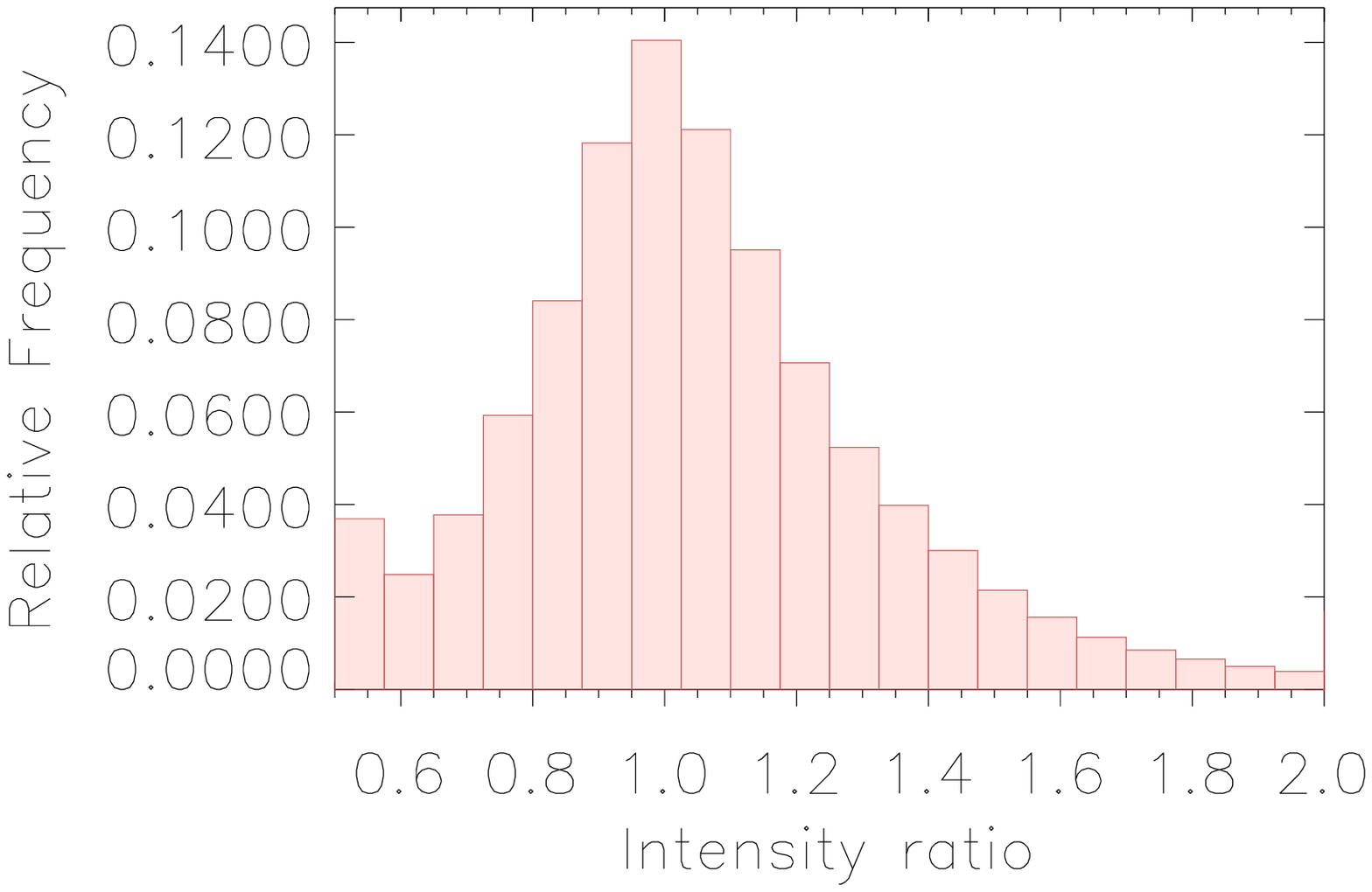}\\
\includegraphics[height=0.28\linewidth]{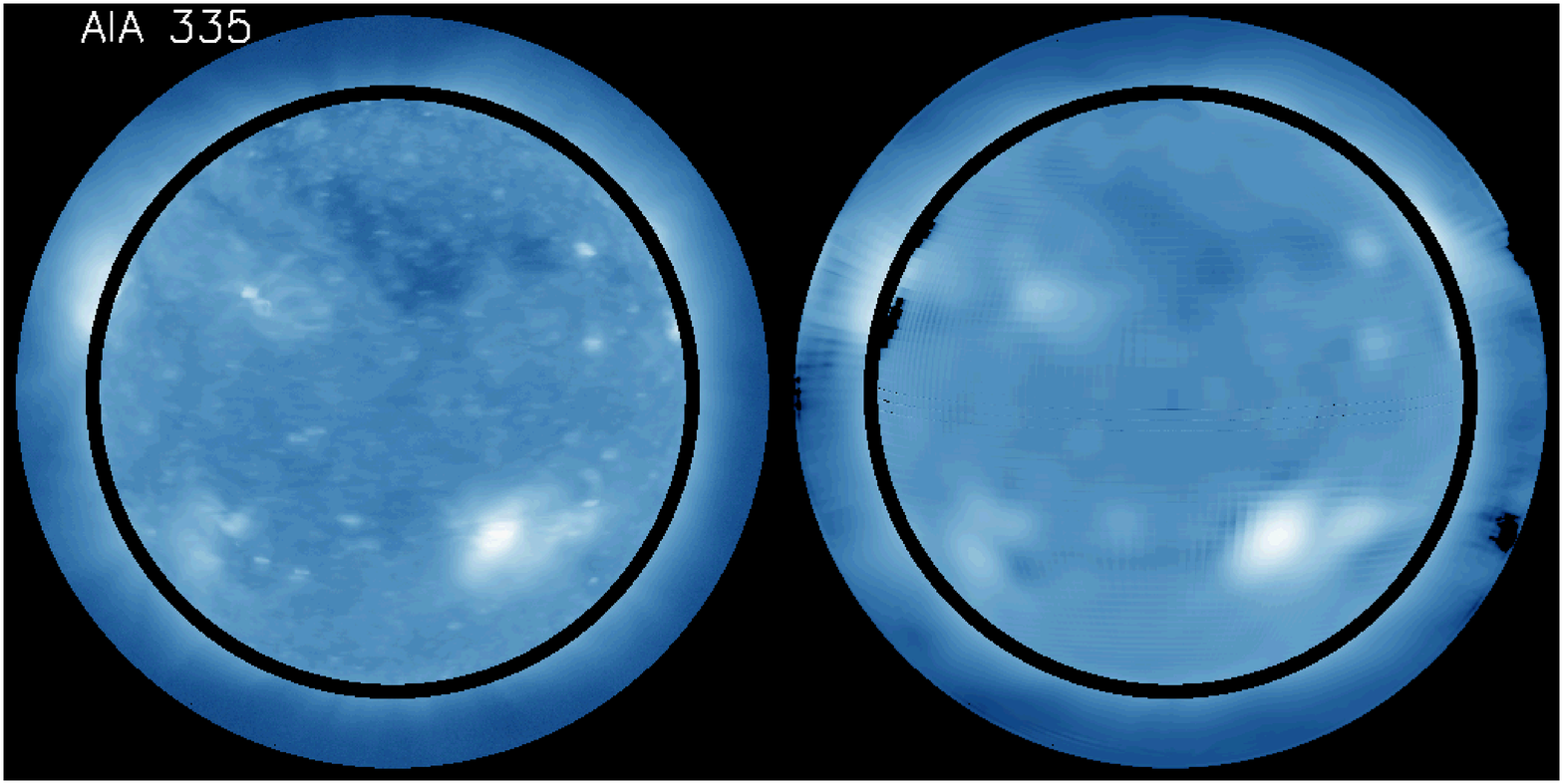}	
\includegraphics[height=0.28\linewidth]{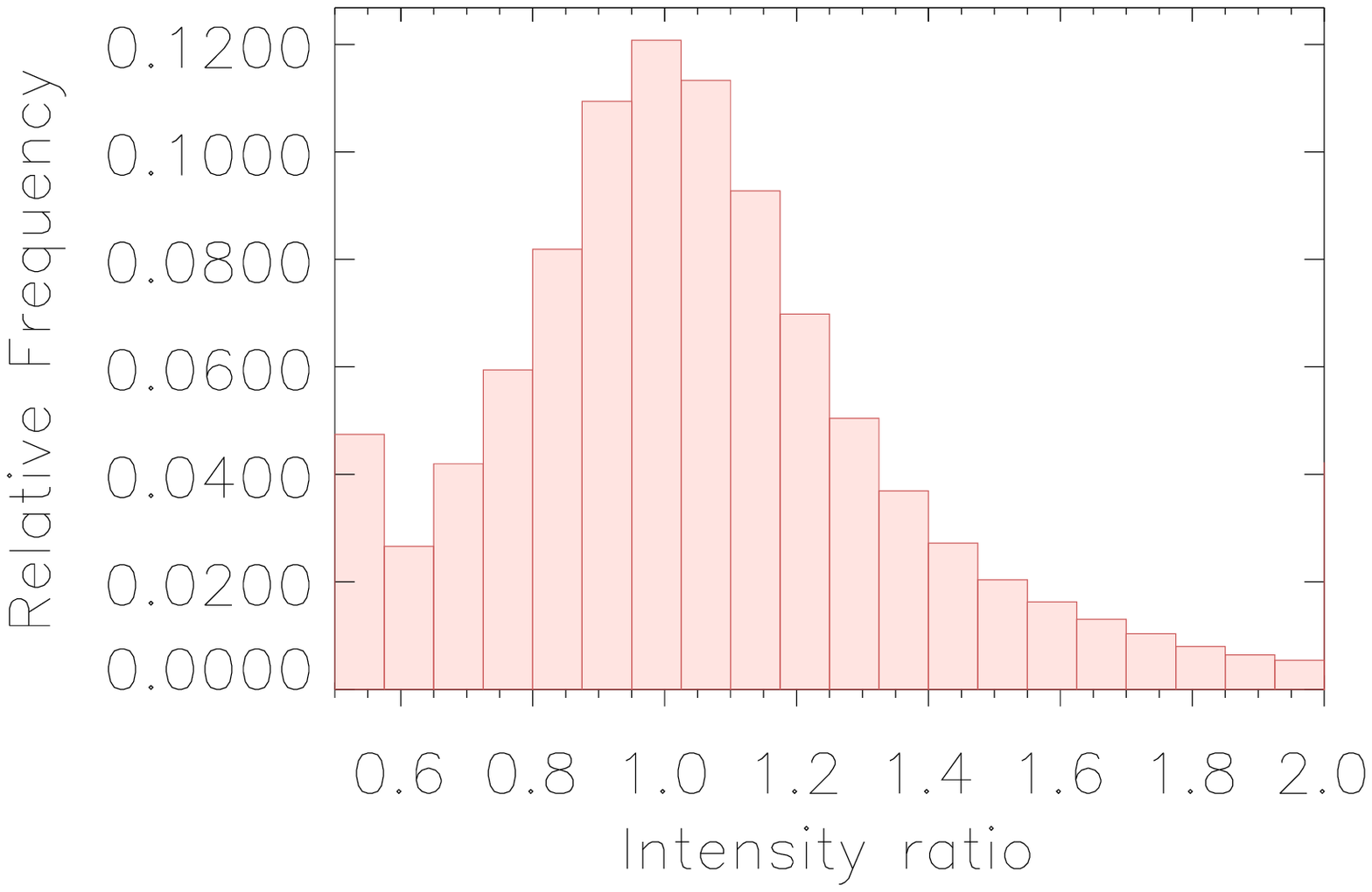}\\
\end{center}
\caption{Comparison of synthetic data derived from the tomographic model against observed data for the bands 171, 193, 211, and 335 \AA\ (from tom to bottom) of the AIA telescope. Left panels: data images and corresponding synthetic images computed by LOS-integration of the tomographic model. Right panel: frequency histogram of the synthetic to observed intensity ratio for every corresponding pair of pixels in the two images.}
\end{figure}

To evaluate the accuracy of the tomographic model, the synthetic images can be quantitatively compared to the corresponding data images. An example of this is shown in Figure \ref{CompareImages} from a tomographic reconstruction of the solar corona for the bands 171, 193, 211, and 335 \AA\ (from tom to bottom) of the AIA telescope. 

For each pair of images, the relative difference between the synthetic and observed values is below 0.1, 0.2 and 0.3 for 34, 59 and 75\% of the pixels, respectively, $\pm 4\%$ depending on the band. The same level of agreement holds for off-limb or on-disk pixels considered separately. The tomographic model provides then a quite detailed reliable description of the average global corona during the reconstructed period. The black rings in the images shown in Figure \ref{CompareImages} correspond to pixels with projected radius in the range 0.98 to 1.025 $\Rsun$. This near-limb data is not actually used for the tomographic inversion, as the emission along their corresponding line-of-sights can affected by optically thick emission \citep{frazin_2009}.

EUV tomography can currently only be applied from only one or two (in the STEREO era) point-of-view. With such limited simultaneous information the temporal resolution of the technique is of the order of half solar rotation (or about two weeks). Of course, this is the most important limitation of the technique, which is then suitable for studying structures that are stable during their observed transit.

\section{The 3D Distribution of the DEM}\label{LDEM}
 
Once the tomographic step is completed, the FBE of all EUV bands is known at each tomographic computational cell. Within each tomographic voxel the plasma is expected to be multi-thermal. The LDEM is a measure of the thermal distribution within the voxel. 

\begin{figure}[!ht] 
\begin{center} 
\includegraphics[width=0.7\linewidth]{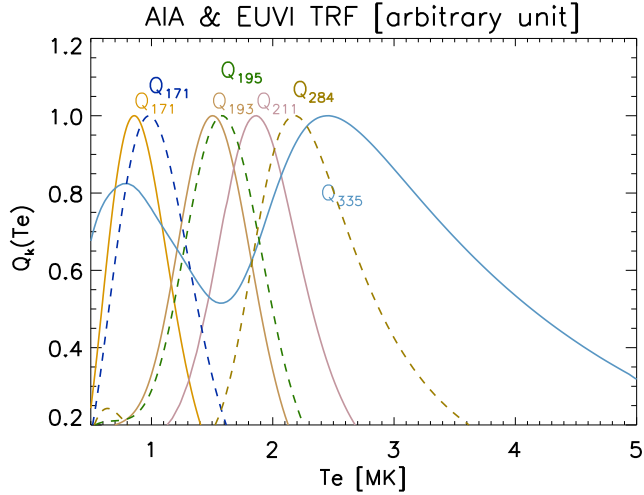}
\end{center}
\caption{Temperature response function of all EUVI coronal bands (dashed, 171, 195, 284 \AA), and 4 AIA coronal bands (solid, 171, 193, 211, 335 \AA). From \citet{nuevo_2015}.} 
\label{fqkl}
\end{figure}

As each FBE represents emission at a different temperature, they provide constraints on the LDEM. Using the passband function of each band of the EUV telescopes, and the atomic database CHIANTI version 7.1 to model of coronal emissivity, Figure \ref{fqkl} shows the \emph{temperature response function} (TRF) of all EUVI and 4 AIA coronal bands. The temperature range modeled by the LDEM is determined by the temperature range of sensitivity of the bands that are used. This can be estimated from the FWHM of the TRFSs around their respective main peaks. For EUVI and the 3 lower temperature bands of AIA, the temperature range is from about 0.5 und up to about 3.0 MK. In the case of AIA, this range is expanded up to $\sim 4$ MK when adding the 335 \AA\ band (see \citet{nuevo_2015}).

Due to the limited number of available data points (FBE values), and also to the narrow band nature of the TRF of each band, the inversion of the LDEM function is under determined and implemented by modeling it with a family of functions depending on a vector of a small number of parameters (typically 3 to 5). When DEMT is based on 3 bands, such as provided by the EUVI telescope, the LDEM is modeled by a single normal function. With the AIA telescope, more coronal bands can be used (up to 6), and the LDEM can be modeled with combinations of normal functions. The reader is referred to \citet{nuevo_2015} for a detailed study on parametric models of the LDEM using both telescopes. 

In each tomographic cell independently, the problem consists of finding the values of the parameter vector to best predict with the LDEM the tomographic values of all FBEs in that cell. To do so an objective function is defined, that measures the quadratic differences between tomographically determined FBEs and the those synthesized from the modeled LDEM.

\begin{figure}[!ht]
\begin{center}
\includegraphics[height=0.29\linewidth]{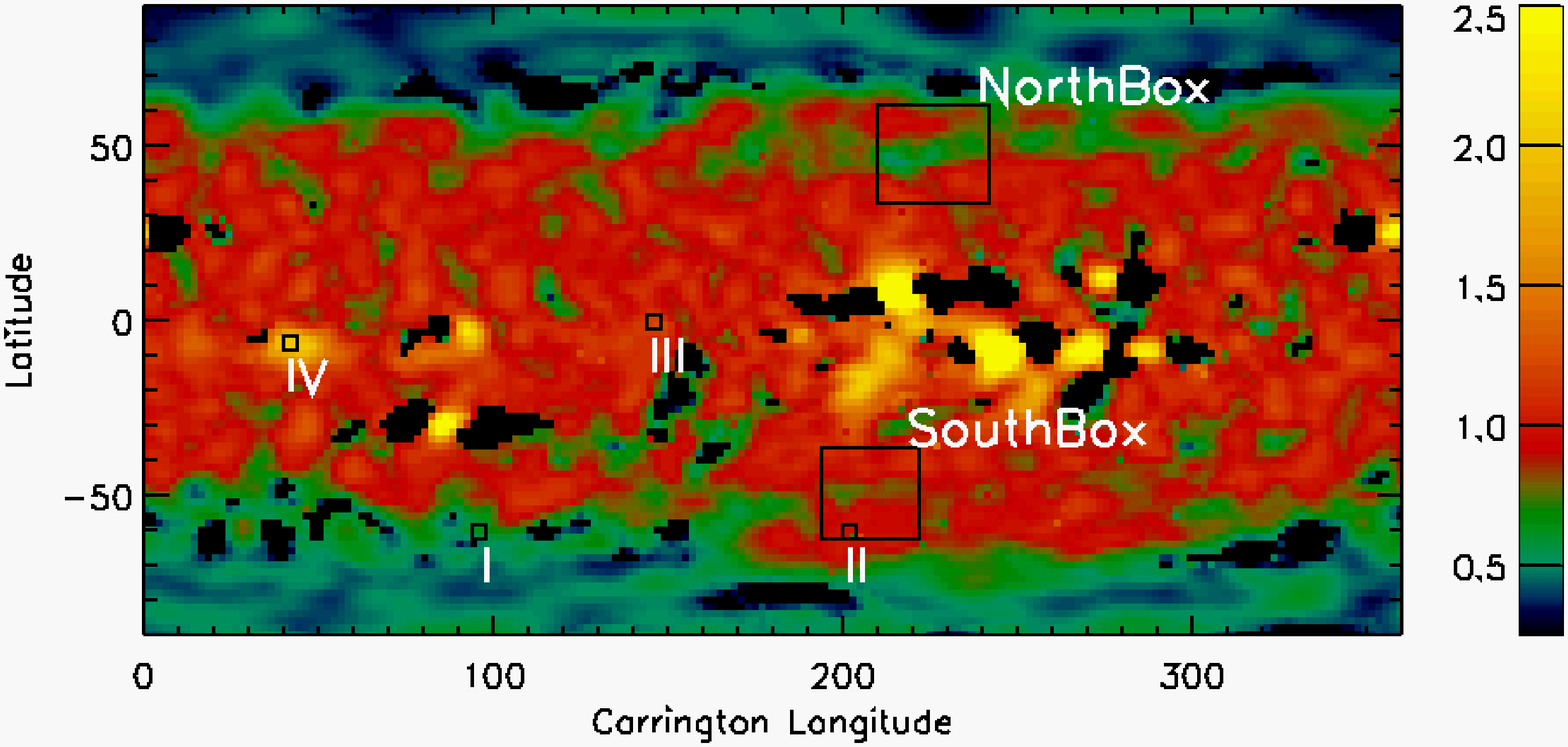}
\includegraphics[height=0.29\linewidth]{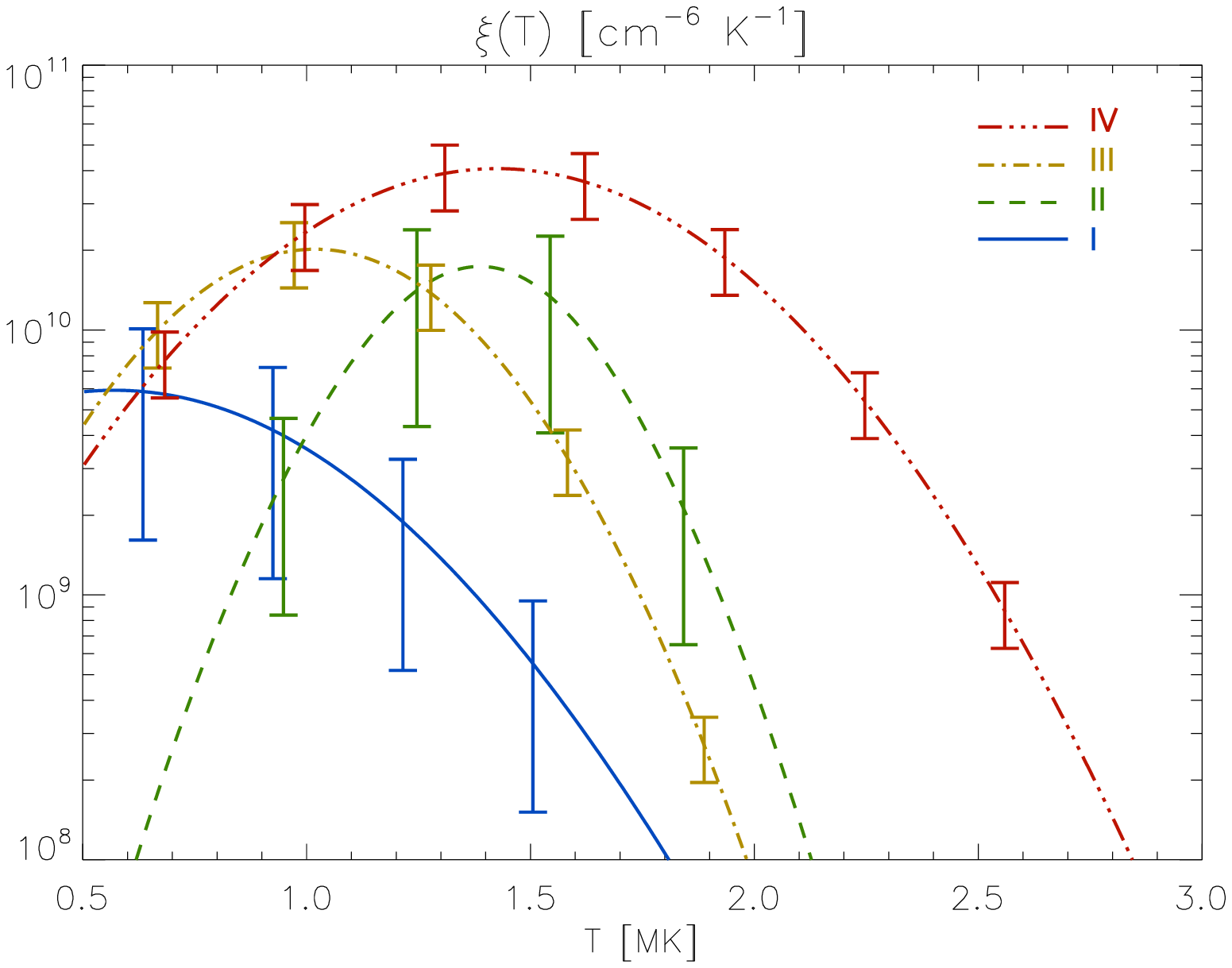}
\end{center}
\caption{\emph{Left:} Latitude-Longitude map of the tomographic electron density $N_e$ in units of $10^8 {\rm cm^{-3}}$, at a height 1.075 R$_\odot$, for the period CR-2069. \emph{Right:} LDEM at the tomographic voxels I, II, III and IV indicated in the left panel. Reproduced from \citet{frazin_2009}.}
\label{DEMT}
\end{figure}

\begin{figure}[!ht]
\begin{center}
\includegraphics[width=0.49\linewidth]{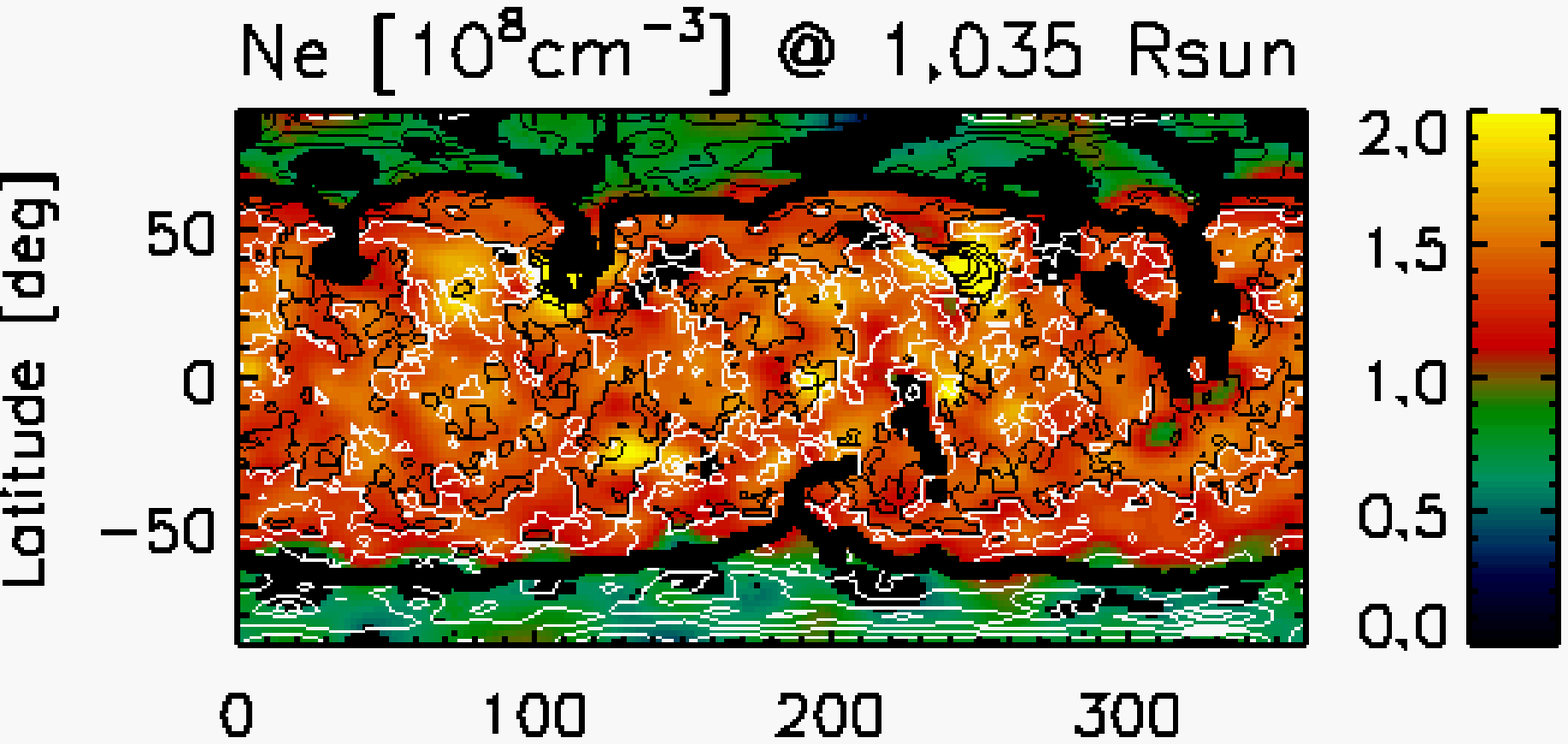}
\includegraphics[width=0.49\linewidth]{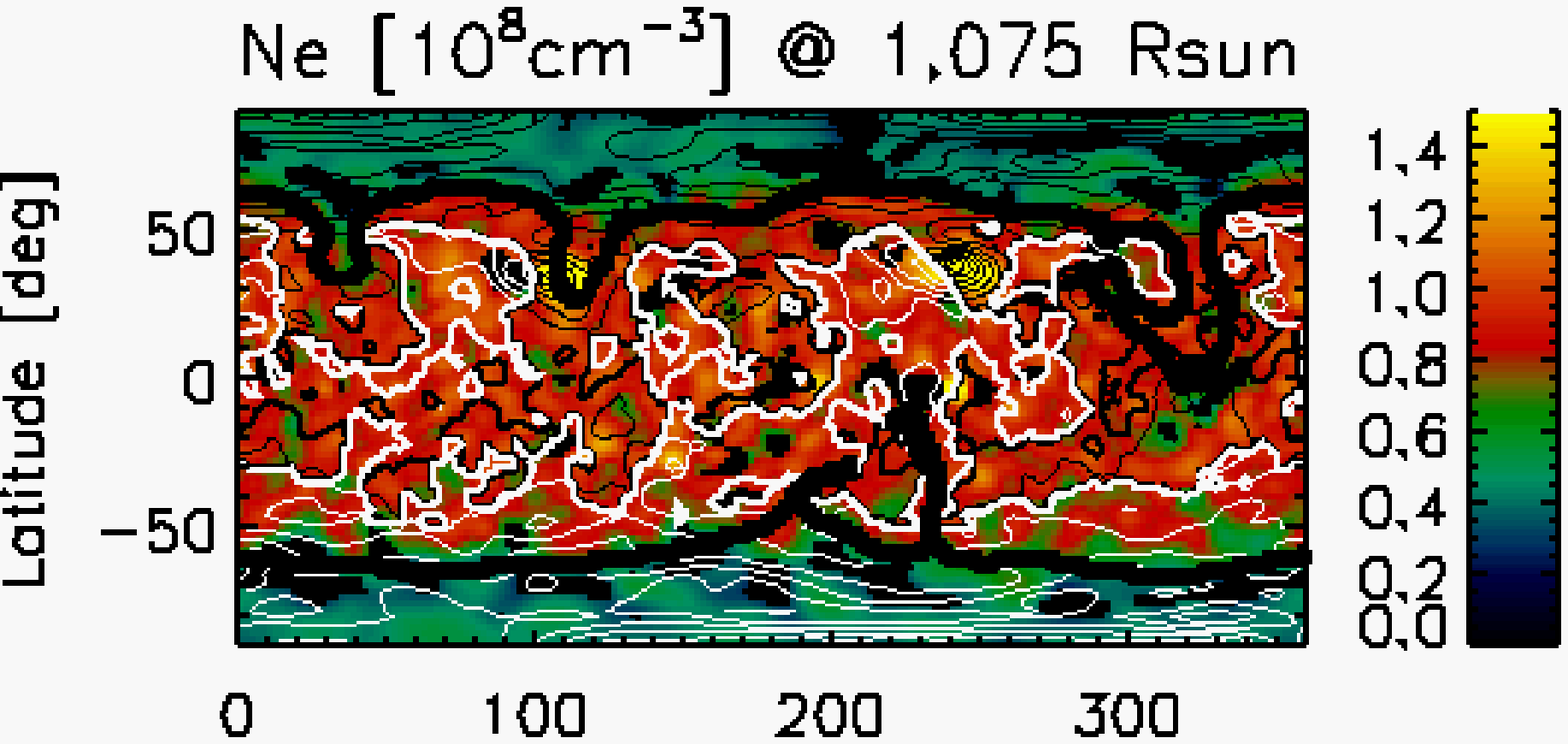}\\
\includegraphics[width=0.49\linewidth]{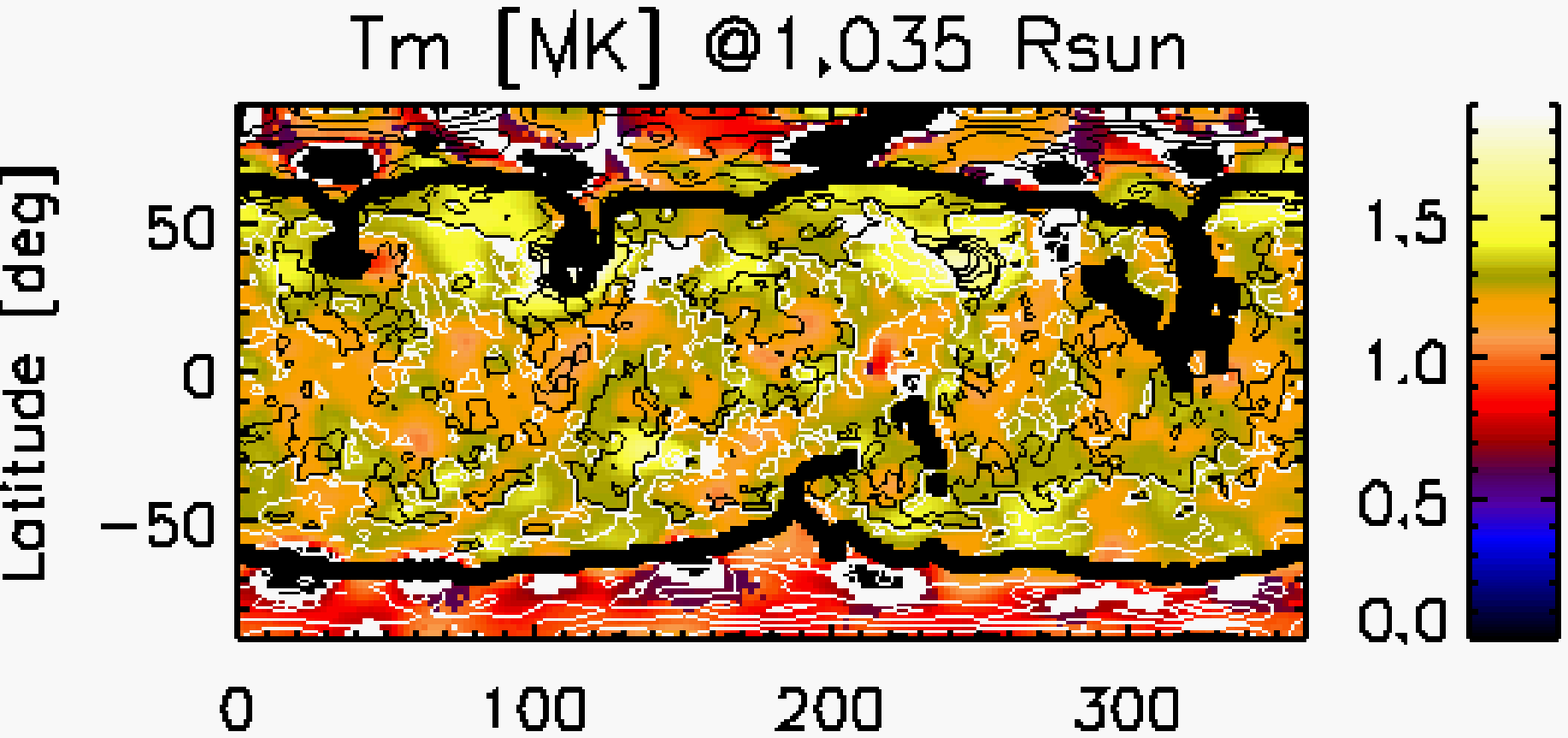}
\includegraphics[width=0.49\linewidth]{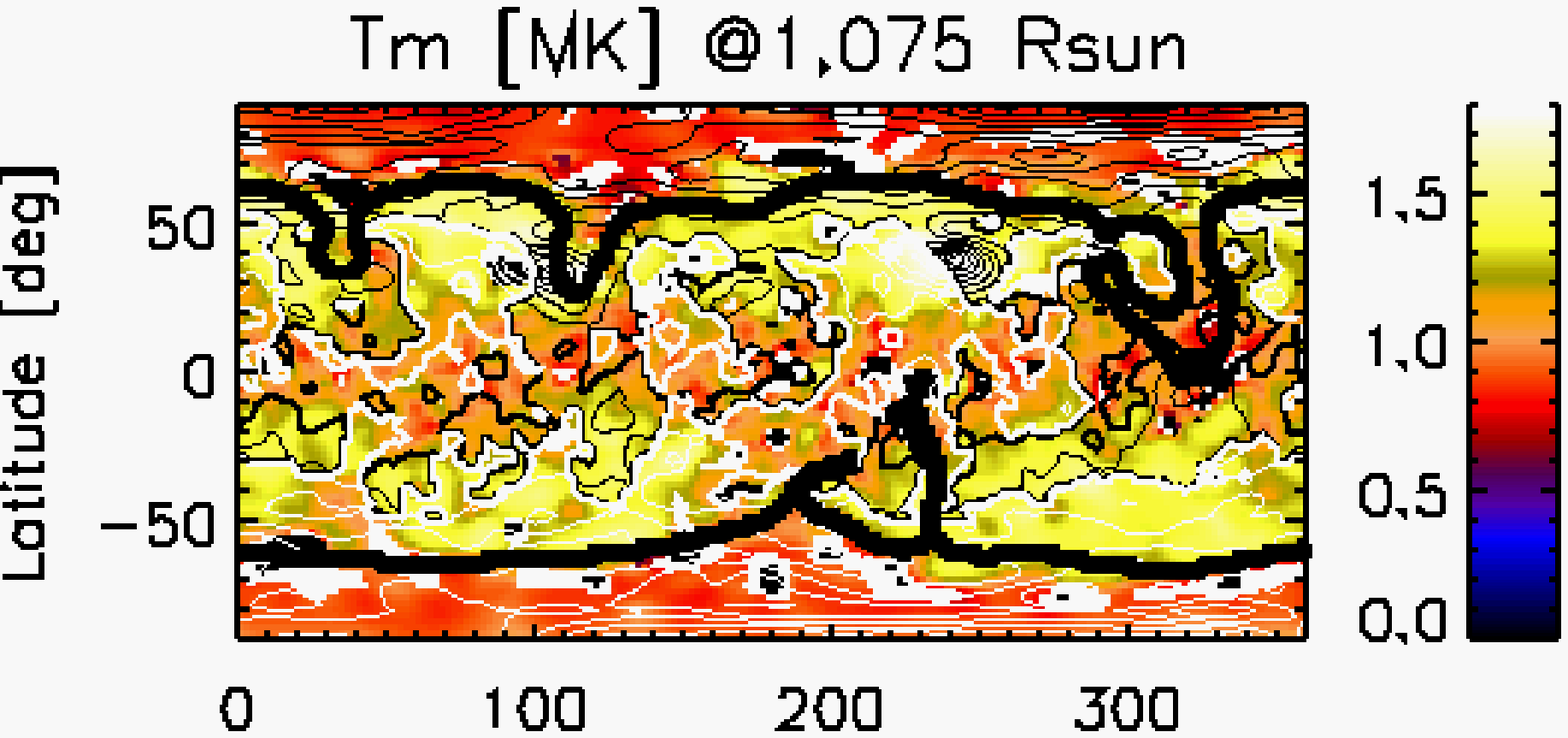}\\
\end{center}
\caption{An example of the 3D reconstruction of the thermodynamical state of the solar minimum corona, during the period CR-2077. Latitude-longitude maps of the electron density (top) and mean electron temperature (bottom) at heights 1.035 (left) and 1.075 $\Rsun$ (right), as derived with the DEMT technique. Solid-thin curves are magnetic strength $B$ contour levels of a PFSS model of the coronal magnetic field, with white (black) representing outward (inward) oriented magnetic field. The solid-thick black curves indicate the location of the magnetically open/closed boundary.}
\label{solmin}
\end{figure}

Once the LDEM is determined, the average electron density in the tomographic cell can be computed from its zeroth moment, specifically as the square root of the integral of the LDEM over temperature. As an example, the left panel in Figure \ref{DEMT} shows a latitude-longitude map of a spherical shell of the electron density obtained at one sample height of the tomographic grid, for Carrington rotation (CR-)2069, a period of very low magnetic activity. At 4 tomographic voxels indicated as I, II, III, and IV in the left panel, the right panel shows the normal LDEM model that is found. The error bars represent the uncertainty due to the regularization level of the tomographic reconstructions. The first moment of the LDEM, divided by its zeroth moment, allows computation of the mean electron temperature predicted by DEMT. Examples of spherical shells of the electron density and mean temperature at two different heights of another tomographic reconstruction are reproduced in Figure \ref{solmin}. Contour levels of the magnetic-strength $B$ of a \emph{potential field source surface} (PFSS) model are over-plotted, along with the magnetically open/closed boundary (see caption in Figure \ref{solmin}).

\section{Results}\label{results}

Applications of DEMT have included the observational study of coronal structures, the use of tomographic results to validate coronal models, and its combination with coronal extrapolations of the photospheric magnetic field. Following we summarize all peer-reviewed published work on DEMT. 

\citet{vasquez_2009} produced the first observational 3D analysis of stable coronal prominence cavities,  measuring the density and temperature contrast between the plasmas in the cavity and in the surrounding helmet streamer. As it is characteristic of tomography, their study did not require any ad-hoc modeling, as needed in the forward modeling approach. 

Being suited to study coronal structures that are stable over half solar rotational time, tomography works best at solar minimum. \citet{vasquez_2010, vasquez_2011} analyzed the global thermodynamical  of the solar corona during the minimum of activity between solar cycles 23 and 24, and discussed their results in relation to the fast and slow components of the solar wind. These works also include comparisons to other observational non-tomographic studies of the same periods, providing cross validation results for the technique. As an example of a 3D reconstruction of the solar minimum, Figure \ref{solmin} displays latitude-longitude maps of the electron density and mean temperature, derived with the DEMT technique for the period CR-2077. At both heights, the location of the open/closed boundary of the PFSS model is characterized by a very high transverse gradient in both the electron density and the mean temperature maps derived from the DEMT analysis.

 \begin{figure}[!ht]
\begin{center}
\includegraphics[height=0.29\linewidth]{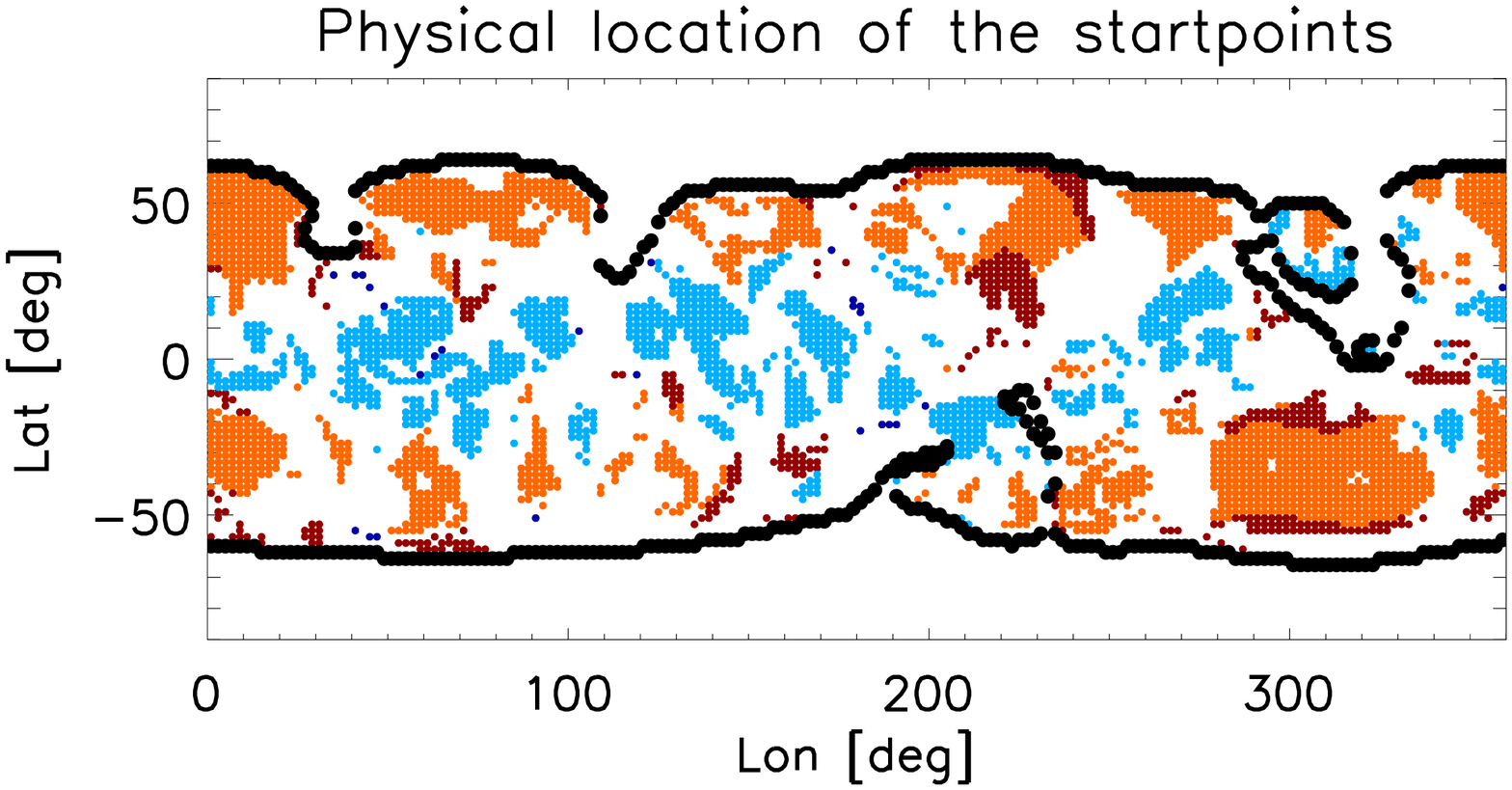}
\includegraphics[height=0.29\linewidth]{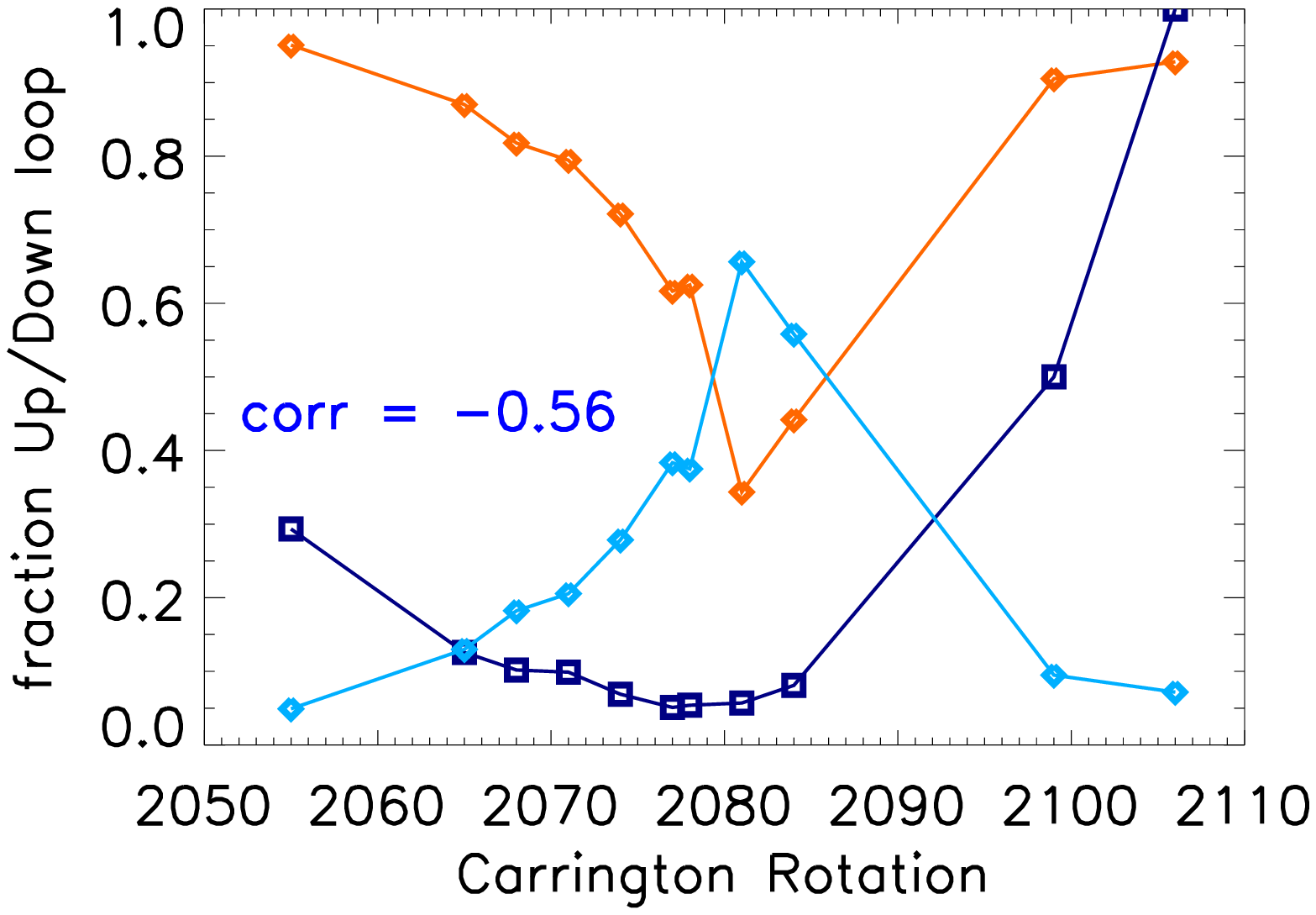}
\end{center}
\caption{\emph{Left:} Location of loops with negative (down) and positive (up) temperature gradient with height, in blue and red, respectively, for CR-2077. \emph{Right:} Evolution of the fraction of down (up) loops indicated as light-blue (red) diamonds, and the sunspot monthly number divided by 32.1 (dark-blue squares). The correlation between the fraction of down loops and the sunspot number is indicated. Adapted from \citet{nuevo_2013}.}
\label{updown}
\end{figure}

DEMT can be combined with the global PFSS magnetic models of the solar corona, an approach dubbed the \emph{Michigan Loop Diagnostic Technique} (MLDT). DEMT results are traced along the field lines of the magnetic model, allowing study of the thermodynamical properties of magnetic flux tubes in the quiet sun in a statistical fashion. \citet{huang_2012} applied MLDT to study one rotation during the last solar minimum. They found the ubiquitous presence of magnetic loops with downward gradients of temperature  dubbed \emph{down loops}. Down loops were found to be dominant in the latitude range $\pm30^\circ$ (left panel in Fig. \ref{updown}). \citet{nuevo_2013} extended the study to a sequence of rotations that included the solar minimum.  Their study revealed a clear anti-correlation between the global coronal activity level and the number of down loops present in the corona (right panel in Fig. \ref{updown}). They found that down and up loops are characterized $\beta\approx 1$ and $\beta<1$, respectively, and proposed an interpretation of their results in terms of Alfv\'en wave damping.

DEMT results have been also used to constrain and validate MHD models. Reconstructions of the electron density and temperature have been used as a constraint to 3D MHD solar wind models coupled to the Space Weather Modeling Framework (SWMF) code suite, that solves for the different electron and proton temperatures \citep{vanderholst_2010}. Improvements on the performance of the models include a more accurate prediction of the occurrence and density of co-rotating interaction regions in the heliosphere. 

\begin{figure}[!ht]
\begin{center}
\includegraphics[width=0.44\linewidth,angle=-90]{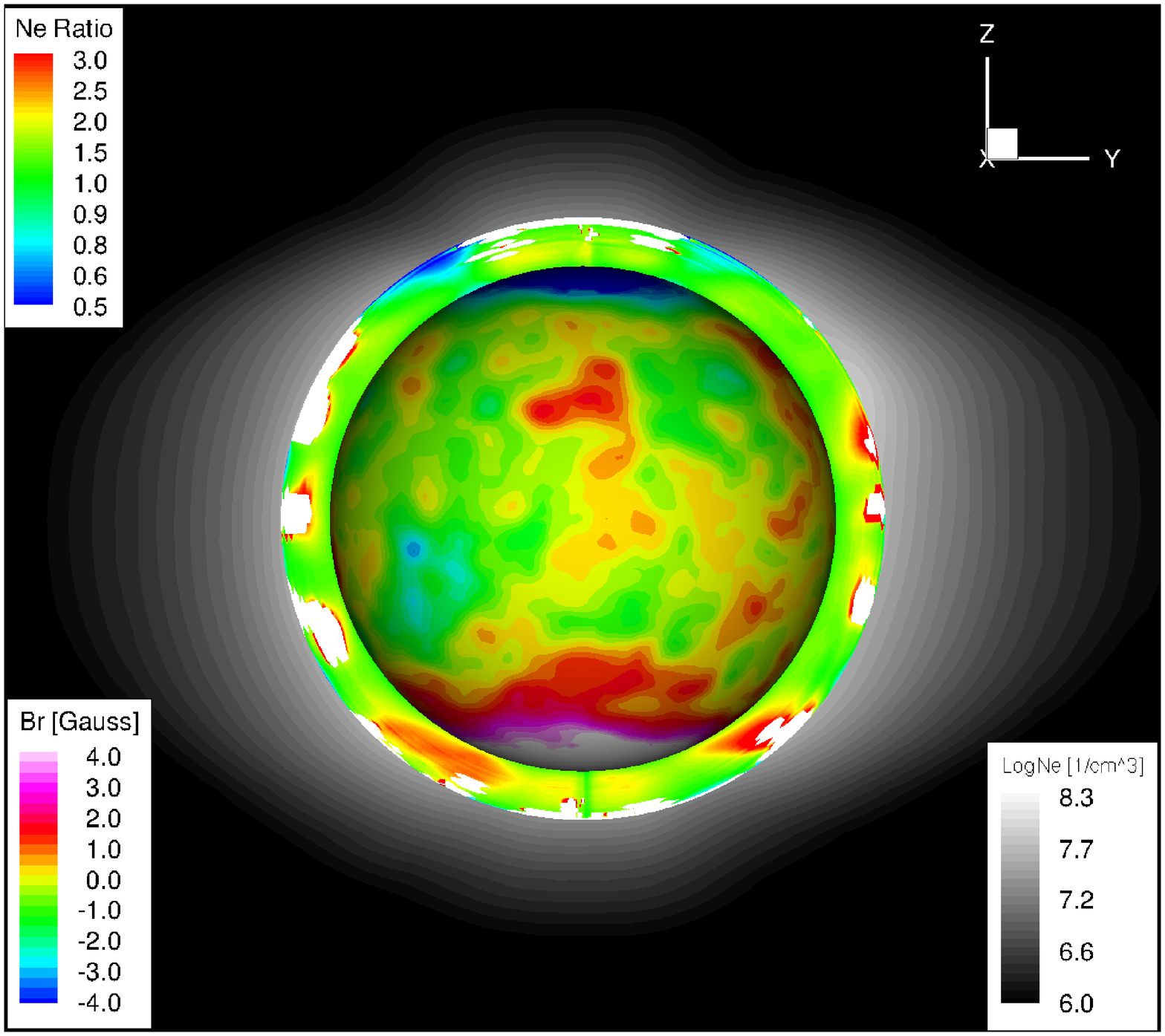}
\includegraphics[width=0.44\linewidth,angle=-90]{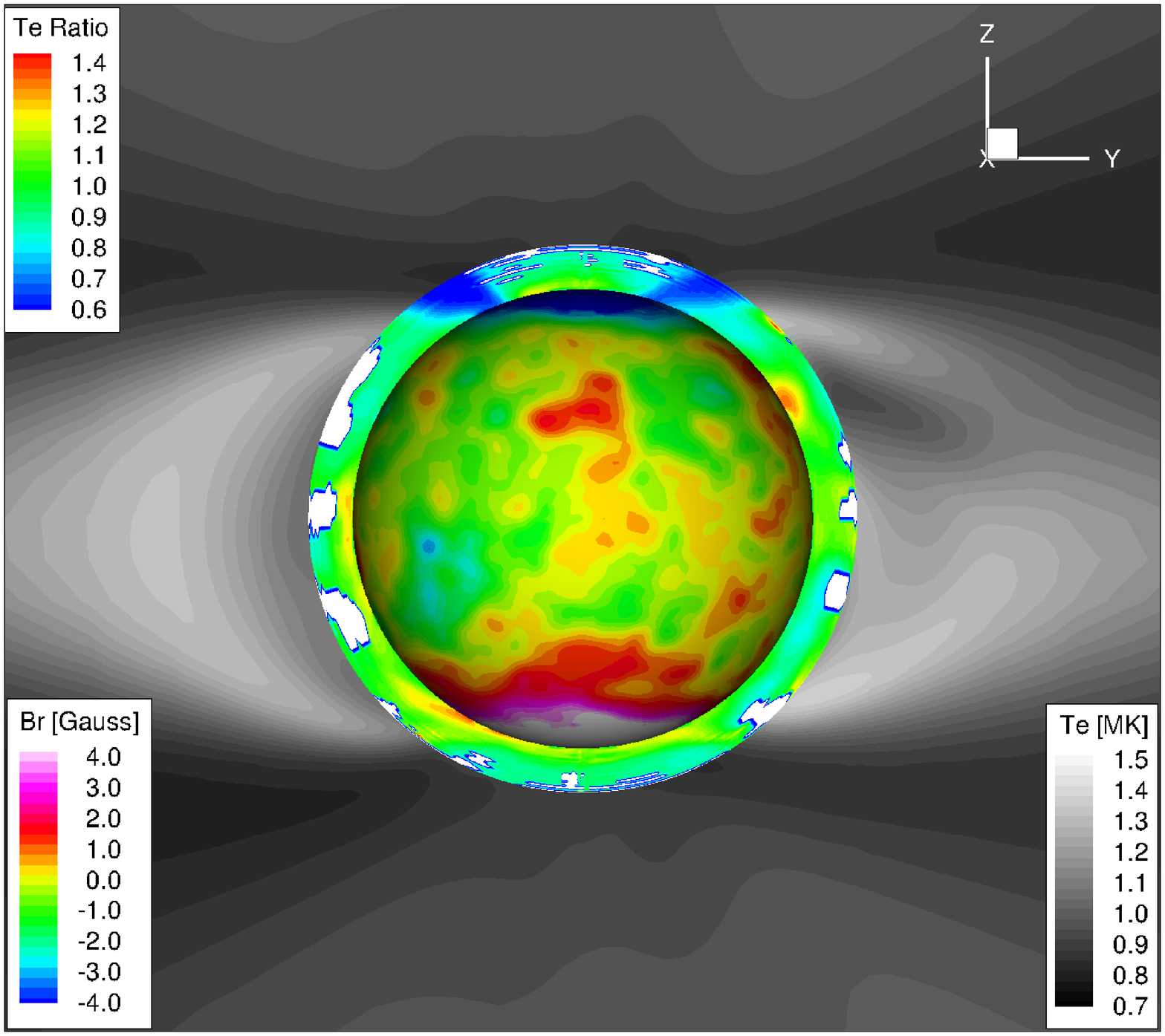}
\end{center}
\caption{DEMT validation of the electron density (left) and temperature (right) of a
two-temperature MHD model of the solar corona, for the period CR-2077. The color rings spanning the height range 1.0 to 1.25 $\Rsun$ show, in a sample meridional cut of the corona, the MHD to DEMT ratio of the respective quantity. The outer greyscale maps ($r> 1.25\, \Rsun$) displays the corresponding MHD results. Reproduced from \citet{jin_2012}.}
\label{validation}
\end{figure}

\begin{figure}[!ht]
\begin{center}
\includegraphics[width=0.7\linewidth]{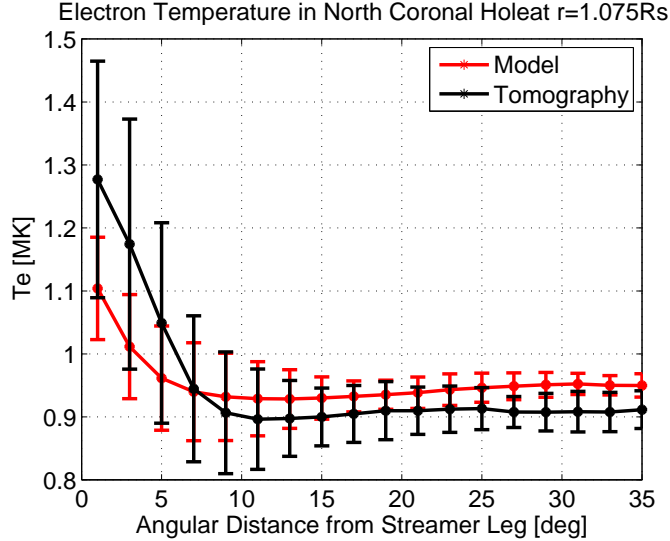}
\end{center}
\caption{DEMT validation of a 3D wave-driven MHD model of the solar wind. 
Electron temperature versus angular distance from the magnetically open/closed boundary (averaged over all longitudes) for the period CR-2063, derived from the MHD model (red) and from the DEMT inversion (black). Reproduced from \citet{oran_2015}.}
\label{validation_2}
\end{figure}

Global two-temperature models of the corona and inner heliosphere have been validated in the inner corona with DEMT reconstructions of the electronic plasma parameters \citep{jin_2012}, with most of the model outputs fitting the observations very well, as seen in Figure \ref{validation}. The 3D products of DEMT have also been used as a validation tool for a study on coronal heating by surface Alfv\'en wave damping, implemented within the MHD model of the solar wind in the SWMF \citep{evans_2012}.  DEMT results have been recently used as a validation tool in a study of the charge state composition of the slow solar wind derived from an ionization evolution code coupled to the wave-driven MHD solar wind model of the SWMF \citep{oran_2015}. DEMT allowed detailed validation of the latitudinal transition of the electronic plasma parameters in the open field lines surrounding the coronal equatorial streamer belt (Figure \ref{validation_2}).

All DEMT studies reviewed so far were based on EUVI data, having 3 EUV bands with a sensitivity range $\sim 0.60 - 2.70$ MK. More recently, \citet{nuevo_2013_b,nuevo_2015} extended the DEMT technique to use the 4 cooler AIA bands (aimed at studying the quiet sun), sensitive to the range $\sim 0.55 - 3.75$ MK. Their study corresponds to CR-2099, a rotation of the rising phase of the current solar cycle 24. While in previous studies the LDEM was always modeled by a single normal distribution, the extra AIA filter, and the increased sensitivity range, allowed exploration of new parametric LDEM models. Using 4 bands, the model that consistently achieves the best predicted-to-reconstructed FBEs is a bimodal distribution, being a superposition of two normal distributions with distinct cool and hot components. The mean centroids of the two components in the quiet diffuse corona are found to be ${\rm log_{10}}\left<T_{0,1}\right>=6.15$ and ${\rm log_{10}}\left<T_{0,2}\right>=6.42$, values that are very consistent with independent determinations of the characteristic temperatures of the solar corona, as discussed in the same study.

\begin{figure}[!ht]
\begin{center}
\includegraphics[width=0.49\textwidth]{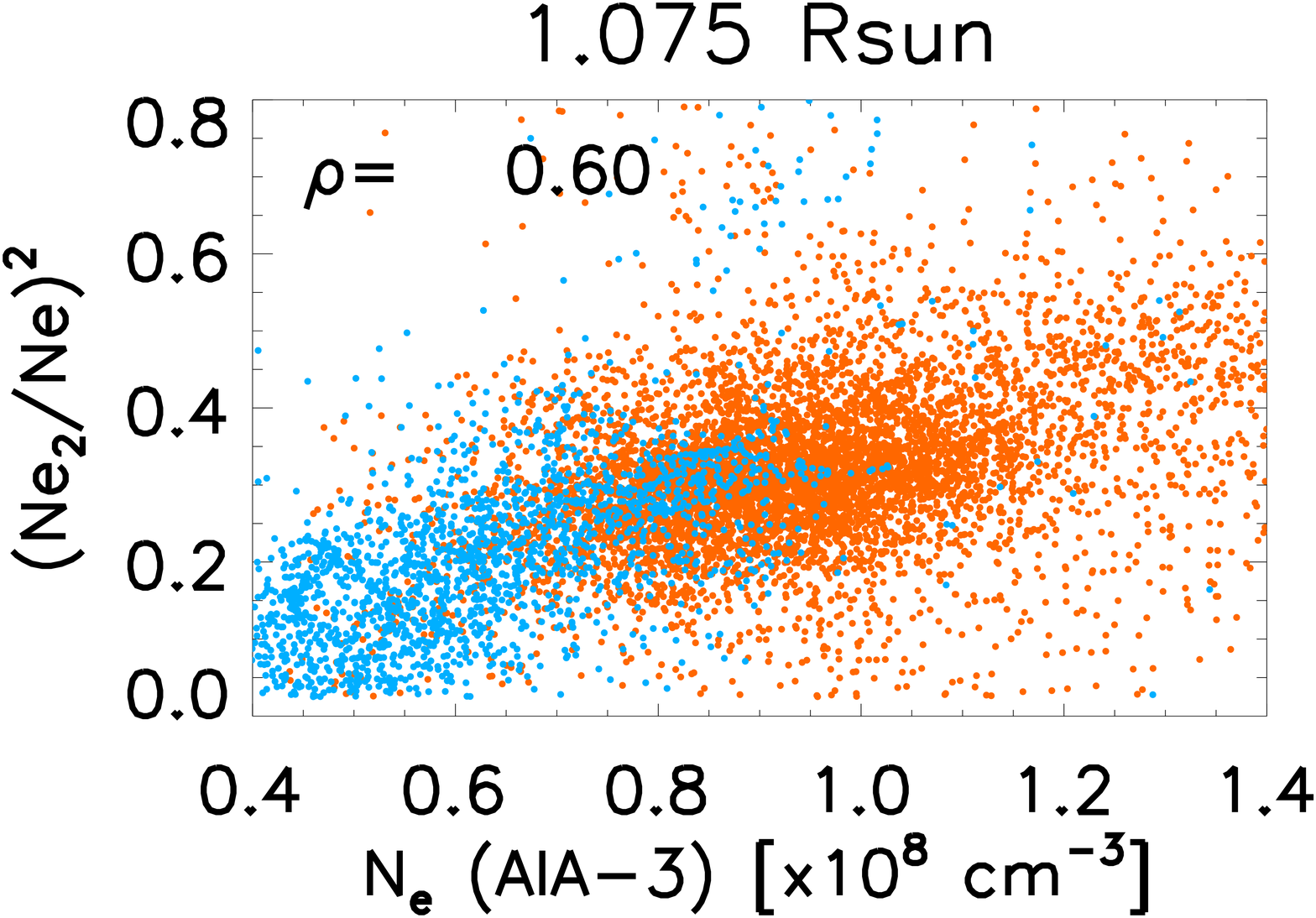}
\includegraphics[width=0.49\linewidth]{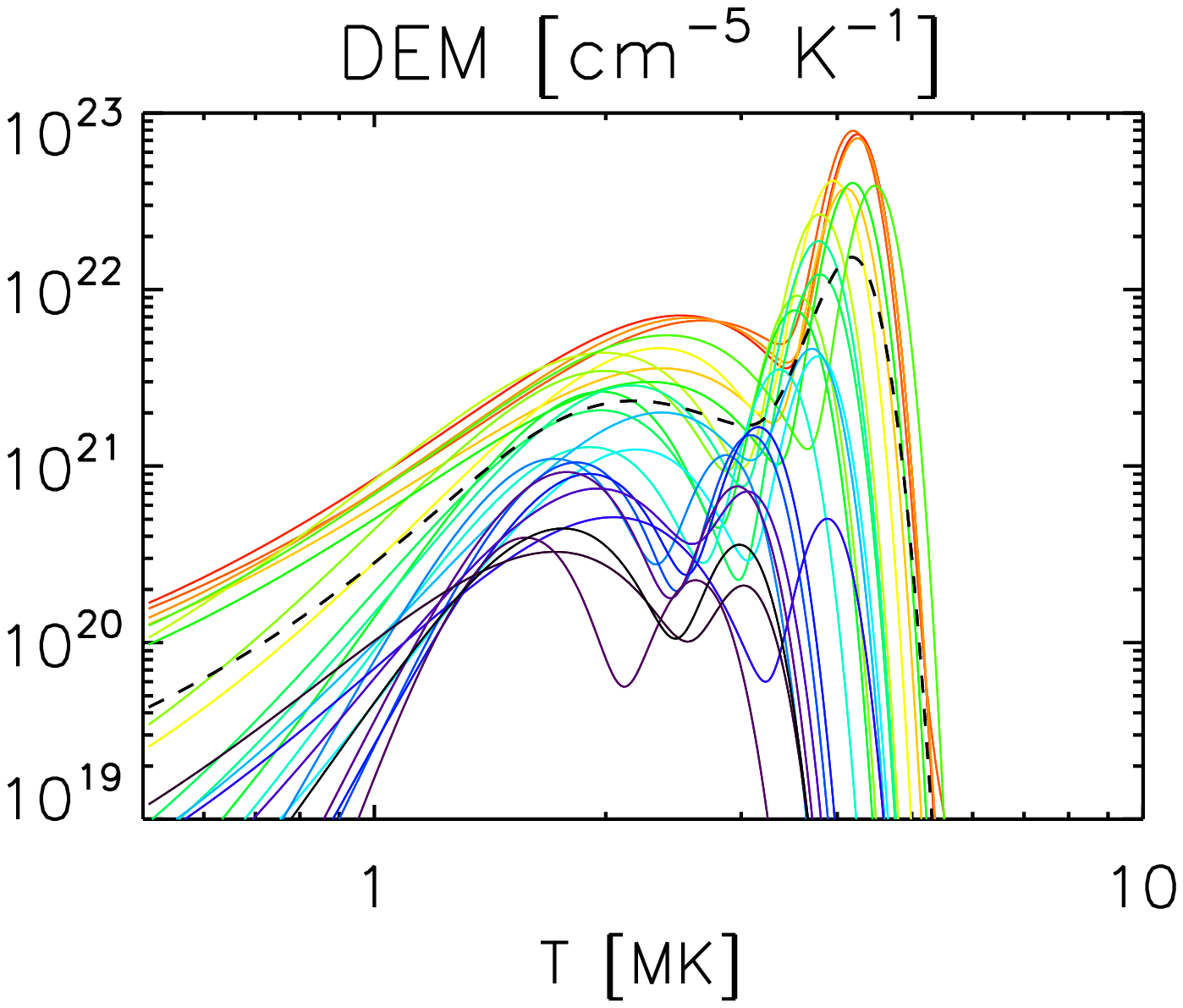}
\end{center}
\caption{\emph{Left:}  Scatter plot of the measure of the bi-modality  $\left(N_{e,2}/N_{e}\right)^2$ (see text) versus $N_e$ derived with AIA, at 1.075 $\Rsun$. \emph{Right:} Bi-modal DEM curves obtained for selected regions within and around an active region. Red/yellow curves correspond to hotter and denser regions, blue/dark curves correspond to colder and less dense regions. Reproduced from \citet{nuevo_2015}.}
\label{bimodal}
\end{figure}

The square electron densities of the two components are $N_{e,1}^2$ and $N_{e,2}^2$, respectively, so that the total square electron density of the LDEM is $N_e^2=N_{e,1}^2+N_{e,1}^2$. A measure of the bi-modality is then the ratio $\left(N_{e,2}/N_{e}\right)^2$. The study of this ratio throughout the diffuse quiet corona reveals that the bimodality of the LDEM is ubiquitous, and that it is stronger for denser regions, as shown by the left panel of Figure \ref{bimodal}.

\citet{nuevo_2015} also validate the LDEM inversion technique by applying it to standard 2D DEM studies. Examples of the bimodal DEMs are shown in the right panel of Figure \ref{bimodal}. The DEMT study shows that LDEM of the quiet corona is bimodal at the spatial resolution of the tomographic grid, which is $0.01 \ \Rsun \times 2^\circ \times 2^\circ$, or about $(7\times10^3 {\rm km}) \times [(2.44\times10^4 {\rm km})^2]$ for a representative voxel at a height of 0.1 $\Rsun$ above the photosphere at the equator.  The authors argue that the nanoflare heating scenario is less likely to explain these results, and that alternative mechanisms, such as wave dissipation appear better supported by them.

\section{Concluding Remarks and Future Prospects}\label{conclusions}

DEMT provides a quantitative average description of the solar corona over a full solar rotation with the following main characteristics,

\begin{itemize}

\item The products are 3D maps of: a) the filter band emissivity in each EUV band, b) the local-DEM, and c) its moments, such as  $\left<N_e^2\right>$ and $\left<T_e\right>$.

\item The temporal resolution of the technique is currently limited to the transit time of coronal features, i.e. $\sim 14$ days. DEMT reconstructions are then reliable descriptions of slowly evolving coronal structures, such as the diffuse quiet corona, and coronal holes. DEMT is not suited to study fastly evolving ARs.

\item The spatial resolution depends on the computational grid size, which in turn is constrained by the cadence of the image series. Typically, one image every 6 hrs is used, a time over which the Sun rotates about $3.3^\circ$. The adopted tomographic cell size is then $2^\circ$ in both angular directions and $10^{-2}\,{\rm R_\odot}$ in the radial direction.

\item The image processing, tomographic inversion, and DEM determination, are fully automated tasks, with little user interaction.

\item Its implementation does not require any ad-hoc modeling.

\item Finite FOV and finite computational grid effects, issues of relevance for white light tomography, are not important for EUV tomography.

\end{itemize}

The current implementation of the DEMT and MLDT techniques involves a suite of codes written in the C and IDL programming languages, which can be efficiently ran in nowadays desktop multi-core computers. Future immediate planned applications of DEMT and MLDT involve comparative studies of the last two solar minima (using the EIT and EUVI instruments), as well as the study of the coronal radiative losses predicted by MLDT in the quiet sun.

Tomographic inversion of time series of full-sun images is the only available observational technique that can provide global constraint and validation to large scale MHD modeling of the corona and the solar wind. As such, DEMT is a highly valuable tool to help in the continued development of global coronal models. DEMT could provide highly valuable 3D maps of the detailed temperature distribution of the coronal plasma if \emph{full-sun} spectral images (of the type provided by the EIS instrument) become available. It would be highly desirable that such an instrument will be operational during the a solar minimum period, when single point-of-view tomography is most fruitful.

The authors thanks CONICET grant PIP IU Nro 11420100100151 to IAFE that has funded this research. The author also thanks the collaboration and useful comments by Federico A. Nuevo and Richard A. Frazin.

\bibliography{amvasquez_bibl}

\begin{thebibliography}{21}
\expandafter\ifx\csname natexlab\endcsname\relax\def\natexlab#1{#1}\fi
\expandafter\ifx\csname url\endcsname\relax
  \def\url#1{\texttt{#1}}\fi
\expandafter\ifx\csname urlprefix\endcsname\relax\def\urlprefix{URL }\fi

\bibitem[{{Altschuler} and {Perry}(1972)}]{altschuler_1972}
{Altschuler}, M.~D., {Perry}, R.~M., Apr. 1972. {On Determining the Electron
  Density Distribution of the Solar Corona from K-Coronameter Data}. \solphys
  23, 410--428.

\bibitem[{{Aschwanden}(2011)}]{aschwanden_2011}
{Aschwanden}, M.~J., Oct. 2011. {Solar Stereoscopy and Tomography}. Living
  Reviews in Solar Physics 8, 5.

\bibitem[{{Evans} et~al.(2012){Evans}, {Opher}, {Oran}, {van der Holst},
  {Sokolov}, {Frazin}, {Gombosi}, and {V{\'a}squez}}]{evans_2012}
{Evans}, R.~M., {Opher}, M., {Oran}, R., {van der Holst}, B., {Sokolov}, I.~V.,
  {Frazin}, R., {Gombosi}, T.~I., {V{\'a}squez}, A., Sep. 2012. {Coronal
  Heating by Surface Alfv{\'e}n Wave Damping: Implementation in a Global
  Magnetohydrodynamics Model of the Solar Wind}. \apj 756, 155.

\bibitem[{{Frazin}(2000)}]{frazin_2000}
{Frazin}, R.~A., Feb. 2000. {Tomography of the Solar Corona. I. A Robust,
  Regularized, Positive Estimation Method}. \apj 530, 1026--1035.

\bibitem[{{Frazin} and {Janzen}(2002)}]{frazin_2002}
{Frazin}, R.~A., {Janzen}, P., May 2002. {Tomography of the Solar Corona. II.
  Robust, Regularized, Positive Estimation of the Three-dimensional Electron
  Density Distribution from LASCO-C2 Polarized White-Light Images}. \apj 570,
  408--422.

\bibitem[{{Frazin} et~al.(2005){Frazin}, {Kamalabadi}, and
  {Weber}}]{frazin_2005}
{Frazin}, R.~A., {Kamalabadi}, F., {Weber}, M.~A., Aug. 2005. {On the
  Combination of Differential Emission Measure Analysis and Rotational
  Tomography for Three-dimensional Solar EUV Imaging}. \apj 628, 1070--1080.

\bibitem[{{Frazin} et~al.(2009){Frazin}, {V{\'a}squez}, and
  {Kamalabadi}}]{frazin_2009}
{Frazin}, R.~A., {V{\'a}squez}, A.~M., {Kamalabadi}, F., Aug. 2009.
  {Quantitative, Three-dimensional Analysis of the Global Corona with
  Multi-spacecraft Differential Emission Measure Tomography}. \apj 701,
  547--560.

\bibitem[{{Huang} et~al.(2012){Huang}, {Frazin}, {Landi}, {Manchester},
  {V{\'a}squez}, and {Gombosi}}]{huang_2012}
{Huang}, Z., {Frazin}, R.~A., {Landi}, E., {Manchester}, W.~B., {V{\'a}squez},
  A.~M., {Gombosi}, T.~I., Aug. 2012. {Newly Discovered Global Temperature
  Structures in the Quiet Sun at Solar Minimum}. \apj 755, 86.

\bibitem[{{Inhester}(2006)}]{inhester_2006}
{Inhester}, B., Dec. 2006. {Stereoscopy basics for the STEREO mission}. ArXiv
  Astrophysics e-prints.

\bibitem[{{Jin} et~al.(2012){Jin}, {Manchester}, {van der Holst}, {Gruesbeck},
  {Frazin}, {Landi}, {Vasquez}, {Lamy}, {Llebaria}, {Fedorov}, {Toth}, and
  {Gombosi}}]{jin_2012}
{Jin}, M., {Manchester}, W.~B., {van der Holst}, B., {Gruesbeck}, J.~R.,
  {Frazin}, R.~A., {Landi}, E., {Vasquez}, A.~M., {Lamy}, P.~L., {Llebaria},
  A., {Fedorov}, A., {Toth}, G., {Gombosi}, T.~I., Jan. 2012. {A Global
  Two-temperature Corona and Inner Heliosphere Model: A Comprehensive
  Validation Study}. \apj 745, 6.

\bibitem[{{Leblanc} et~al.(1970){Leblanc}, {Leroy}, and
  {Poulain}}]{leblanc_1970}
{Leblanc}, Y., {Leroy}, J.~L., {Poulain}, P., May 1970. {The Characteristic
  Sizes and the Electron Density of Coronal Enhancements Observed in White
  Light}. \aap 5, 391.

\bibitem[{{Minnaert}(1930)}]{minnaert_1930}
{Minnaert}, M., 1930. {On the continuous spectrum of the corona and its
  polarisation. With 3 figures. (Received July 30, 1930)}. \zap 1, 209.

\bibitem[{{Nuevo} et~al.(2013{\natexlab{a}}){Nuevo}, {Huang}, {Frazin},
  {Manchester}, {Jin}, and {V{\'a}squez}}]{nuevo_2013}
{Nuevo}, F.~A., {Huang}, Z., {Frazin}, R., {Manchester}, iv, W.~B., {Jin}, M.,
  {V{\'a}squez}, A.~M., Aug. 2013{\natexlab{a}}. {Evolution of the Global
  Temperature Structure of the Solar Corona during the Minimum between Solar
  Cycles 23 and 24}. \apj 773, 9.

\bibitem[{{Nuevo} et~al.(2013{\natexlab{b}}){Nuevo}, {V{\'a}squez}, {Frazin},
  and {Landi}}]{nuevo_2013_b}
{Nuevo}, F.~A., {V{\'a}squez}, A.~M., {Frazin}, R.~A., {Landi}, E.,
  2013{\natexlab{b}}. {Multi-modal DEM in the solar corona}. Boletin de la
  Asociacion Argentina de Astronomia La Plata Argentina 56, 395--398.

\bibitem[{{Nuevo} et~al.(2015){Nuevo}, {V{\'a}squez}, {Landi}, and
  {Frazin}}]{nuevo_2015}
{Nuevo}, F.~A., {V{\'a}squez}, A.~M., {Landi}, E., {Frazin}, R., 2015.
  {Multimodal Differential Emission Measure in the Solar Corona}. \apj,
  submitted.

\bibitem[{{Oran} et~al.(2014){Oran}, {Landi}, {van der Holst}, {Lepri},
  {V{\'a}squez}, {Nuevo}, {Frazin}, {Manchester}, {Sokolov}, and
  {Gombosi}}]{oran_2015}
{Oran}, R., {Landi}, E., {van der Holst}, B., {Lepri}, S.~T., {V{\'a}squez},
  A.~M., {Nuevo}, F.~A., {Frazin}, R., {Manchester}, IV, W.~B., {Sokolov},
  I.~V., {Gombosi}, T.~I., Dec. 2014. {A Steady-State Picture of Solar Wind
  Acceleration and Charge State Composition Derived from a Global Wave-Driven
  MHD Model}. ArXiv e-prints.

\bibitem[{{van de Hulst}(1950)}]{vandehulst_1950}
{van de Hulst}, H.~C., Feb. 1950. {The electron density of the solar corona}.
  \bain 11, 135.

\bibitem[{{van der Holst} et~al.(2010){van der Holst}, {Manchester}, {Frazin},
  {V{\'a}squez}, {T{\'o}th}, and {Gombosi}}]{vanderholst_2010}
{van der Holst}, B., {Manchester}, IV, W.~B., {Frazin}, R.~A., {V{\'a}squez},
  A.~M., {T{\'o}th}, G., {Gombosi}, T.~I., Dec. 2010. {A Data-driven,
  Two-temperature Solar Wind Model with Alfv{\'e}n Waves}. \apj 725,
  1373--1383.

\bibitem[{{V{\'a}squez} et~al.(2009){V{\'a}squez}, {Frazin}, and
  {Kamalabadi}}]{vasquez_2009}
{V{\'a}squez}, A.~M., {Frazin}, R.~A., {Kamalabadi}, F., May 2009. {3D
  Temperatures and Densities of the Solar Corona via Multi-Spacecraft EUV
  Tomography: Analysis of Prominence Cavities}. \solphys 256, 73--85.

\bibitem[{{V{\'a}squez} et~al.(2010){V{\'a}squez}, {Frazin}, and
  {Manchester}}]{vasquez_2010}
{V{\'a}squez}, A.~M., {Frazin}, R.~A., {Manchester}, IV, W.~B., Jun. 2010. {The
  Solar Minimum Corona from Differential Emission Measure Tomography}. \apj
  715, 1352--1365.

\bibitem[{{V{\'a}squez} et~al.(2011){V{\'a}squez}, {Huang}, {Manchester}, and
  {Frazin}}]{vasquez_2011}
{V{\'a}squez}, A.~M., {Huang}, Z., {Manchester}, W.~B., {Frazin}, R.~A., Dec.
  2011. {The WHI Corona from Differential Emission Measure Tomography}.
  \solphys 274, 259--284.

\end{thebibliography}
\bibliographystyle{elsarticle-harv}

\end{document}